\documentclass[11pt]{article}
\usepackage{setspace}
\usepackage{graphicx}
\usepackage[space]{grffile}
\usepackage{amsmath}
\usepackage{amsthm}
\usepackage[sort,comma,authoryear]{natbib}
\usepackage{mathtools}
\usepackage{bm}
\usepackage{amssymb}
\usepackage{threeparttable}
\usepackage{hyperref}

\usepackage{booktabs}
\usepackage{array}
\usepackage[english]{babel}
\usepackage{longtable}
\usepackage{subfig}
\usepackage{url}
\usepackage{ragged2e}
\usepackage{tabularx}
\usepackage{epstopdf}
\usepackage[toc,title,page]{appendix}

\usepackage{longtable}
\usepackage{threeparttable}
\usepackage{rotating}

\theoremstyle{definition}

\theoremstyle{remark}
\newtheorem{remark}{Remark}
\usepackage{tabularx}
\usepackage{changepage}
\usepackage{verbatim}
\usepackage[a4paper, total={6in, 8in}]{geometry}
\usepackage{setspace}
\usepackage{booktabs}
\usepackage{geometry}
\usepackage{multirow}
\numberwithin{equation}{section}
\usepackage{geometry}
\usepackage{ulem}

\newcommand{\tabincell}[2]{\begin{tabular}{@{}#1@{}}#2\end{tabular}}
\usepackage{hyperref}
\usepackage{natbib}
\hypersetup{
	colorlinks,
	linkcolor={blue},
	citecolor={blue},
	urlcolor={blue}
}
\usepackage[ruled]{algorithm2e}
\begin{document}

%\linenumbers

\date{}

\title{Dynamic online prediction model and its application to automobile claim frequency data}
%\title{Modelling automobile claim frequency with dynamic online prediction model}

\author{
	Jiakun Jiang \footnote{Center for Statistics and Data Science, Beijing Normal University, Zhuhai, China.}
	\quad  Zhengxiao Li \footnote{School of Insurance and Economics, University of International Business and Economics, Beijing, China. Email: li\_zhengxiao@uibe.edu.cn.}
	\quad Liang Yang \footnote{corresponding author: School of Finance, Southwestern University of Finance and Economics, Chengdu, China. Email: yangliang@swufe.edu.cn.}
}

\maketitle

\begin{abstract}
Prediction modelling of claim frequency is an important task for pricing and risk management in
non-life insurance and needed to be updated frequently with the changes in the insured population, regulatory legislation and technology.
Existing methods are either done in an ad hoc fashion, such as parametric model calibration, or less so for the purpose of prediction.
In this paper, we develop a Dynamic Poisson state space (DPSS) model which can continuously update the parameters whenever new claim information becomes available. DPSS model allows for both time-varying and time-invariant coefficients.  To account for smoothness trends of time-varying coefficients over time, smoothing splines are used to model time-varying coefficients. 
The smoothing parameters are objectively chosen by maximum likelihood. The model is updated using batch data accumulated at pre-specified time intervals, which allows for a better approximation of the underlying Poisson density function.
The proposed method can be also extended to the distributional assumption of zero-inflated Poisson and negative binomial.
In the simulation, we show that the new model has significantly higher prediction accuracy compared to existing methods. We apply this methodology to a real-world automobile insurance claim data set in China over a period of six years and demonstrate its superiority by comparing it with the results of competing models from the literature.
\\
\\
%	{\bf{JEL Classification Numbers:}}
{\bf{Article History:}}  \today
\end{abstract}
\newpage

\section{Introduction}
Determining future numbers of claims
is one of the main tasks that actuaries perform in the insurance industry, especially in the insurance ratemaking process. {A well-performed prediction model should maintain its accuracy over time.}
Most existing prediction methods in actuarial science can be featured by static prediction models, such as generalized linear models and their extensions \citep{klein2015bayesian, lee2021addressing, lambert1992zero, yip2005modeling}.
Many insurance examples have shown that the performance of static prediction models easily tends to worsen over time.
This is usually due to changes over time in the populations of insured groups, irregular claims behaviors, legislative change, or technological progress \citep{zhu2011modeling, chang2020dynamic, dong2013, HENCKAERTS2022}.
Some methods have been proposed to address this issue, such as model refitting and recalibration. However, these methods are often done in an ad hoc fashion. In this paper, we propose a dynamic prediction model in which the parameter estimates and prediction of future numbers of claims can be sequentially updated over time. It is well suited for massive data streams in which the data-generating process is itself changing over time. Furthermore, it is an online implementation that allows rapid updating of model parameters as new claim data arrive.

The frequently used methodology of predicting future claim frequency based on multi-year insurance claim data is
the credibility theory, in which the time dependence structure is modelled by random effects \citep{denuit2021wishart, lee2020poisson}.
Recently, several credibility models with dynamic random effects have been adapted
to a longitudinal context to conduct experience ratemaking,
where the risk of a policyholder is governed by an individual process of unobserved heterogeneity
\citep{taylor2018evolutionary, pinquet2020poisson, Jae2022, Lu2018}.
However, it remains some challenges in the insurance dynamic ratemaking literature.
Firstly, the fact that the prediction models mentioned above have to rely on longitudinal data is a very strong limitation, as the changing population of insured groups every year
makes it difficult for insurers to collect and collate individual policies into longitudinal data.
These approaches will result in the reduction of massive claim information due to the exclusion of early surrendered and non-renewed policies.
Secondly, insures will be plagued by the computational burden for predicting future premiums in the dynamic random effect
models as the Bayesian credibility premium typically lacks tractability \citep{LiH2021}.
Thirdly, while the existing methods allow random effects to be varied over time \citep{pinquet2020poisson},
they are still static models where the regression parameters are assumed
fixed throughout the data, which will cause
the model's parameter estimates to
track the parameters poorly when the effect of risk factors is subject to variation over time \citep{chang2020dynamic}.
Moreover, these models are too restrictive for capturing complicated nonlinear relationships between response variables and covariates over time.
These motivate us to
propose a relatively simple prediction model structure for numbers of claim as functions of several important rating factors,
but allow the model parameters to vary with time and capture the complicated nonlinear effects of time.
This leads our study of varying-coefficient models \citep{hastie1993varying, fan2008statistical}.
Specifically, for this context,
since the varying-coefficient models are readily interpretable, we can provide a standard interface between the statistical model and the insurance ratemaking mechanism by reporting the fitted individual regression parameters.
However, the vary-coefficient models in statistics
mainly focus on estimation and statistical inference rather than future prediction, as both kernel and spline-based estimators are not suitable for prediction.

State space models are powerful tools in modelling dynamic systems and have been developed for both Gaussian and non-Gaussian response, with its application in biology \citep{bhattacharjee2018bayesian},
medicine \citep{jiang2021dynamic},
epidemiology \citep{o2002tutorial} and insurance \citep{lai2012evolutionary, Jae2022, fung2017}.
In this method, the dynamic system is modelled over time by a set of latent parameters that determine the changing distribution of observed data. The hierarchical model structure allows for both making statistical inferences and predicting future outcomes, taking into account the observed past trajectory. For Gaussian distribution, there is a closed-form solution for estimating the parameters. For the non-Gaussian distribution, such as the claim count data in our motivating example, the solution often relies on complicated numerical integration Normal approximations like the Laplace approximation are commonly used \citep{Lewis1997} for computational convenience.

%The varying-coefficient regression models
%have been very popular in statistics literature in recent years, in which the coefficients of GLMs are replaced by smoothing non-parametric functions and hence the regression coefficients are allowed to vary as functions of other factors \citep{hastie1993varying}.
%A seminal review of the varying-coefficient models given by \cite{fan2008statistical} discusses three estimation methods, including
%kernel smoothing, polynomial splines, and smoothing splines, and also discusses the applications of the varying-coefficient models in various contexts, including
%survival analysis, times series data modelling and so on.
%While one could use non-parametric models to provide more flexibility, they often suffer from the ``curse of dimensionality" \citep{park2015varying}.
%To overcome this problem, \cite{wang2014boosted} develop a boosted varying-coefficient regression model using regression trees as the base learner that is scalable to high-dimensional data.
%\cite{chatla2020tree} propose a tree-based semi-varying coefficient model for the Conway-Maxwell-Poisson distribution which is a two-parameter generalization of the Poisson distribution and is flexible enough to capture both under-dispersion and over-dispersion in count data.
%\blue{Those existing literature in statistics mainly focused on estimation and statistical inference rather than the future  prediction. Both kernel and spline based estimators are not suitable for prediction.
%}

In insurance applications, the system characteristics change slowly over time and might be subject to changes in insurance regulatory legislation. A commonly used method is to model the parameter of interest as a smooth function over time, for example, smoothing spline.
\cite{Wahba1978} and \cite{wahba1983bayesian} show that the estimation of a smoothing spline for a Gaussian outcome is equivalent to the posterior estimate of a Gaussian stochastic process.
\cite{gu1992penalized} shows that it also holds for Non-Gaussian outcomes with the Laplace approximation.
\cite{wecker1983signal}
show how to fit smoothing splines using the state space model formulation.
{Recently, \cite{jiang2021dynamic} discuss the dynamic binary classification problem based on state space model for clinical decision marking.
However, 
this approach is not suitable for insurance ratemaking as the claim count data usually exhibits the over-dispersion features with an excess of zeros.}
Motivated by modelling the claim frequency of automobile insurance based on the multi-year claim data,
we propose a Dynamic Poisson state space (DPSS) model in which time-varying coefficients modelled by the state equations.
The underlying distribution assumption of Poisson distribution can be extended to negative-binomial distribution and
zero-inflated Poisson distribution to capture the zero-inflated and over-dispersion features of claim count data.
%\footnote{The Dynamic negative-binomial state space model (DNBSS) and Dynamic zero-inflated Poisson state space model (DZIPSS) are also discussed in this paper}.
Similar to the traditional Poisson model, the DPSS model regards the number of claims as the response variable, and employs a log-link function to connect the response variable
to the covariates. In contrast to the traditional Poisson model, the time-varying coefficients in the DPSS model are modelled using smoothing splines, similar to \cite{wecker1983signal}.
%Similar to \cite{wecker1983signal}, the time-varying coefficients are modelled using smoothing splines.
The smoothing parameters in the proposed models are estimated
via maximum likelihood on the training data.
The K-steps ahead prediction for claim frequency is based on an algorithm similar to
Kalman filter, which allows for the smoothing parameters,  model coefficients, and also prediction to be updated dynamically when new claim data becomes available. 
%The proposed model is designed for 
The advantage of the proposed method is that the updates only need to be computed on the additional observations when new claim data becomes available, whereas it is only updated as each new observation becomes available in the standard state space model. 
Since the filtering step often involves a Laplace approximation for count outcomes, updating one observation at a time will lead to large cumulative approximation errors.
In addition, it is often not feasible or desirable to update the model one data at a time in applications. Consequently, we propose to update the model at pre-specified fixed time intervals when a batch of claims data becomes available.
%when the claim count data exhibits the over-dispersion features with an excess of zeros \citep{yip2005modeling}.

Although there has been a branch of actuarial science literature on state space models in mortality forecasting \citep{chang2020dynamic} and non-life insurance ratemaking \citep{Jae2022},
there are few studies to discuss how to embed state space models into varying-coefficient models for future prediction.
To the best of our knowledge, this is the first study
of the dynamic prediction model under a claim-frequency-modelling framework.
By adopting the time-varying coefficient model with tools of state-space modelling, our DPSS model significantly fills the gap in dynamic claim frequency modelling and forecasting.
The computational efficiency of our proposed method enables it to be implemented for online monitoring.
It should also be noted that while the proposed method is designed for predicting claim frequency, it can be easily extended to the prediction of claim severity and pure premium by assuming the response follows other continuous heavy-tailed distribution, i.e, gamma distribution, or semi-continuous distribution, i.e.,
compound Poisson distribution or Tweedie distribution \citep{Jorgensen1994}.

To demonstrate the application of the proposed method, we analyze a novel Chinese data set on automobile insurance, followed over the years from 2010 to 2015 with claim history and self-reported risk characteristics of approximately 440,000 policies.
Our goal is to start from a ratemaking model with static parameters and to develop a dynamic mechanism that adjusts the baseline price of rating factors utilizing the introduction of time-varying coefficients.
Strong empirical evidence
suggests this model's superiority over the static prediction model.
Moreover, we present empirical evidence of the time-varying effect, including
\textit{expo}, \textit{carusage}, \textit{carseats}, \textit{carorigin},
\textit{coverage}, \textit{gender}, \textit{ptype} and \textit{svolume} as
estimated by the DPSS model \footnote{The description of these covariates can be found in Tabel \ref{tab-claim-factors} }.
The decreasing trend after the year 2012 can be observed for the estimated regression coefficients of most of the rating factors.
We argue that these dynamic patterns can be caused by the changing automobile regulatory policy in China as regulators tighten insurers' rights to set their insurance premiums, aiming to promote uniform automobile premium rate standards.
%Simulation evidence is further provided to support this argument.
Therefore, identifying and understanding the time-varying effects and the causes of these effects
is critical for theoretical research and actuarial ratemaking practice.

The structure of this paper is as follows.
In Section \ref{sec: model-specifications} we
first provide a summary of the partial varying coefficient model and state space model
and then introduce a new class of Dynamic Poisson state space models.
Estimation and prediction procedure are discussed in Section \ref{sec: estimation}.
The advantages of DPSS model compared to several established methods in the insurance ratemaking literature
are illustrated by a simulation study in Section \ref{sec: simulation}.
To illustrate its practical use, in Section \ref{sec: applications}, we conduct an analysis of the DPSS model fitted to a Chinese automobile insurance data set over most six years periods.
Section \ref{section:conclusion} gives some conclusions and future possible extensions.

\section{Methodology}\label{sec: model-specifications}
\subsection{Partial varying coefficient model}
Classical generalized linear models are not feasible to capture the time-varying effects of covariates. A more sophisticated model is varying-coefficient models \citep{hastie1993varying}.  They allow the regression coefficients to vary in a smooth way with another variable (for example time).
Let $(y_i, \bm{x}_i, t_i)$ for $i=1,\cdots,n$ be the observed data for each subject $i$ at observed time $t_i$,
where $0\leq t_1 < t_2 <\cdots < t_n \leq 1$,
$\bm{x}_i=(x_{i1},\cdots,x_{iq})^T$ denotes a $q$-dimensional vector of covariates, and $y_i$ is the insured claim count (the number of claims).
We propose the following model for $y_i$ in which $\lambda_i$ is the intensity (expected claim frequency), that is,
\[
y_i\sim \text{Poisson}(\lambda_i),
\]
with
\begin{align}
\log \left( {{\lambda }_{i}} \right)&=\beta_0(t_i)+\sum_{j=1}^{q_1}\beta_j(t_i)x_{ij} + \sum_{k=1}^{q_2}\alpha_k x_{i,q_1+k} \nonumber \\
&=\bm{x}_{i1}^{T}\bm{\beta} \left( {{t}_{i}} \right)+\bm{x}_{i2}^{T}\bm{\alpha},
\label{eq-DPSM-1}
\end{align}
where $\bm{x}_{i1}=(1,x_{i1},\cdots,x_{iq_1})^T$ and $\bm{x}_{i2}=(x_{i,q_1+1},\cdots,x_{i,q})^T$
are the covariates
with varying coefficients $\bm{\beta}(t)=(\beta_0(t),\beta_1(t),\cdots,\beta_{q_1}(t))^T$
and with time-invariant coefficients $\bm{\alpha}=(\alpha_1,\cdots,\alpha_{q_2})^T$ respectively, and $q_1+q_2=q$. The varying-coefficient regression models have got lot of researches in statistics literature since proposed by \cite{hastie1993varying}; see \cite{fan2008statistical, park2015varying} for a comprehensive review. However, most existing literatures are all focused on the problem of estimation and statistical inference rather than prediction.
%A seminal review of the varying-coefficient models given by \cite{fan2008statistical}
\begin{remark}
In the context of automobile insurance ratemaking,
we usually assumes that the number of claims, $Y_i$, is Poisson distribution with mean $\mathbb{E}(Y_i)=w_i \lambda_i$, where
$w_i$ is risk exposure and $\lambda_i = \exp(\bm{x}_i^T\bm{\beta})$. Then we have $\log \mathbb{E}(Y_i)=\log(w_i)+\bm{x}_i^T\bm{\beta}$, in which $\log(w_i)$ can be viewed as covariate with fixed coefficient 1.
In (\ref{eq-DPSM-1}) we treat $\log(w_i)$ as covariates with unknown coefficients.
Thus, the traditional Poisson regression model can be regarded as a special case of \eqref{eq-DPSM-1}.% In addition, the interaction terms defined as the product of covariates can be included in the model.
\end{remark}

\begin{remark}
Although Poisson distribution is a natural candidate for modelling count data, it is commonly found that the number of zeros
in insurance claim data exceeds the number of zeros predicted
from the Poisson model.
To model the excessive zeros phenomenon in claim count distribution, the zero-inflated Poisson (ZIP) model proposed by \cite{lambert1992zero} is usually used to fit
automobile insurance claims \citep{yip2005modeling}.
To deal with overdispersion, where the assumption of equal expectation and variance inherent in the Poisson distribution has to be replaced by variances exceeding the
expectation, the negative-binomial distribution provides a convenient
framework extending the Poisson distribution by a second parameter determining
the scale of the distribution, see for example \cite{hilbe2011negative}.
Using the well-known negative-binomial (NB) and zero-inflated Poisson (ZIP), we further extend the Dynamic Poisson state space model (DPSS) to Dynamic negative-binomial state space model (DNBSS) and Dynamic zero-inflated Poisson state space model (DZIPSS). The model specifications and estimation procedures can be found in appendix \ref{app-DZPSS} and \ref{app-DNBPSS} for more details.
\end{remark}

The parameters can be estimated through maximizing
the following penalized log-likelihood function:
\begin{align}\label{eq-log-likelihood}
{{L}_{c}}\left(\bm{\alpha} ,\bm{\beta}(t) ; \bm{y} \right)&=\sum\limits_{i=1}^{n}{\left\{ {{y}_{i}}\left[ \bm{x}_{i1}^{T}\bm{\beta} \left( {{t}} \right)+\bm{x}_{i2}^{T}\bm{\alpha}  \right]-\exp \left[ \bm{x}_{i1}^{T}\bm{\beta} \left( {{t}_{i}} \right)+\bm{x}_{i2}^{T}\bm{\alpha}  \right]	-\log \left( {{y}_{i}}! \right) \right\}}\nonumber \\
& \quad \quad
-\frac{1}{2}\sum\limits_{j=0}^{{{q}_{1}}}{{{\tau }_{j}}\int{{{\left[ {{{{\beta }''}}_{j}}\left( t \right) \right]}^{2}}dt}},
\end{align}
where the $\tau_j$ is smoothing parameter.
It is well known that the minimizer ${\hat{\beta}}_j(t)$ for $j=0, 1, \cdots, q_1$ are cubic smoothing splines.
For data from exponential families, \cite{gu1992penalized} establishes a connection between the smoothing spline models and a Bayesian model from which the estimate of a smoothing spline can be obtained through the posterior estimate of a Gaussian stochastic process. Following Gu's Bayesian approach, $\beta_{j}(t)$ is modelled as
\begin{equation}\label{eq-beta-bayesian}
{{\beta }_{j}}\left( t \right)={{B}_{1j}}+{{B}_{2j}}t+b_{j}^{1/2}\int_{0}^{t}{{{W}_{j}}\left( s \right)ds},\ \ j=0,\cdots,{{q}},
\end{equation}
where $B_{1j}$ and $B_{2j}$ have diffuse prior $\left[B_{1j}, B_{2j}\right]^T \sim N(0, \tau I)$, with $\tau \to \infty$, and $W_j(s)$ are Wiener processes and $b_j$ are smoothing parameters.
The mean of the approximate posterior
is the smoothing spline estimator $\hat{\beta}_j(t)$ when $b_j=1/\tau_{j}$ under diffuse prior, if one approximates the posterior via
Laplace's method \citep{gu1992penalized}.
\cite{wahba1983bayesian} further showed that the stochastic model \eqref{eq-beta-bayesian} can be written with the function $\beta_j(t_i)$ and its first derivative $\beta^\prime_j(t_i)$ in the state vector as:
\begin{align}
\begin{bmatrix}
\beta_j(t_i)\\
\beta_j'(t_i)
\end{bmatrix}
&=T_i\begin{bmatrix}
\beta_j(t_{i-1})\\
\beta_j'(t_{i-1})
\end{bmatrix} + U_j(t_i, t_{i-1}),\nonumber\\
T_i& =\begin{bmatrix}
1& t_{i} - t_{i-1}\\
0 & 1
\end{bmatrix},
\end{align}
where
\begin{align}
U_j(t_i, t_{i-1})&\sim N\left(\begin{bmatrix}
0\\0
\end{bmatrix}, \tau_j^{-1}Q_i
\right),\  Q_i = \begin{bmatrix}
(t_{i} - t_{i-1})^3/3 & (t_{i} - t_{i-1})^2/2 \\
(t_{i} - t_{i-1})^2/2 & (t_{i} - t_{i-1})
\end{bmatrix}.
\end{align}

The parameter $\tau_j$ is the smoothing parameter that controls
the trade-off between smoothness and bias. When $\tau_j^{-1}=0$,  $\beta_j(t)$ is restricted to be a straight line with intercept $\beta_j(0)$
and slope $\beta_j'(0)$. Thus the model \eqref{eq-DPSM-1} can be represented in
a state space representation as
\begin{align}
\log \left( {{\lambda }_{i}} \right)&=\bm{z}_{i}^{T}\bm{\gamma} \left( {{t}_{i}} \right),\ \ \ \ i=1,\cdots ,n  \nonumber\\
\bm{\gamma} \left( {{t}_{i}} \right)&={\bm{T}_{i}}\bm{\gamma} \left( {{t}_{i}} \right)+\bm{\eta} \left( {{t}_{i}},{{t}_{i-1}} \right),\ \ \bm{\eta} \left( {{t}_{i}},{{t}_{i-1}} \right)\sim N\left(\bm{0},{\bm{Q}_{i}} \right) ,
\label{eq-DPSM-2}
\end{align}
where
\begin{align*}
{\bm{z}_{i}}&={{\left( 1,0,{{x}_{i1}},0,{{x}_{i2}},\cdots ,{{x}_{i{{q}_{1}}}},0,{{x}_{i,{{q}_{1}}+1}},\cdots ,{{x}_{iq}} \right)}^{T}}, \\
\bm{\gamma} \left( t \right)&={{\left( {{\beta }_{0}}\left( t \right),{{{{\beta }'}}_{0}}\left( t \right),{{\beta }_{1}}\left( t \right),{{{{\beta }'}}_{1}}\left( t \right),\cdots ,{{\beta }_{{{q}_{1}}}}\left( t \right),{{{{\beta }'}}_{{{q}_{1}}}}\left( t \right),{{\alpha }_{1}},\cdots ,{{\alpha }_{{{q}_{2}}}} \right)}^{T}}, \\
\bm{\eta} \left( {{t}_{i}},{{t}_{i-1}} \right)&={{\left( U_{0}^{T}\left( {{t}_{i}},{{t}_{i-1}} \right),\cdots ,U_{{{q}_{1}}}^{T}\left( {{t}_{i}},{{t}_{i-1}} \right) \right)}^{T}},\\
{\bm{T}}&=\text{diag}\left( \overbrace{{{T}},\cdots ,{{T}}}^{{{q}_{1}}+1},\overbrace{1,\cdots ,1}^{{{q}_{2}}} \right), \\
{\bm{Q}}&=\text{diag}\left( \overbrace{\lambda _{0}^{-1}{{Q}},\cdots ,\lambda _{{{q}_{1}}}^{-1}{{Q}}}^{{{q}_{1}}+1},\overbrace{0,\cdots ,0}^{{{q}_{2}}} \right).
\end{align*}
Here, we refer the model \eqref{eq-DPSM-2} to Dynamic Poisson state space (DPSS) prediction model.
The distinctions between model specifications for varying and constant coefficients are reflected in both the
transition matrix $\bm{T}_i$ and variance matrix $\bm{Q}_i$.
The identity transition in $\bm{T}_i$ that corresponds to $\bm{\alpha}$ and zero variance in $\bm{Q}_i$ implies that $\bm{\alpha}$ are constant coefficients.
In state-space models, the estimates of the coefficients are
updated as each new observation becomes available. For
non-Gaussian state space models, numerical integrations
or normal approximations are typically used in the sequential updates.
Numerical approximation of the count distribution has a large approximation error that can accumulate over time.  In addition, it is often not feasible or desirable to update the model one sample at a time in real applications.
Instead, we propose to update the estimates
over fixed time intervals with batch data, which can lead to more accurate approximation.
%In
%state space models, the estimates of the coefficients are
%updated as each new observation becomes available. For
%non-Gaussian state space models, numerical integrations
%or normal approximations are typically used in the sequential updates.
%Numerical approximation of the binary distribution has a large approximation error that can accumulate over time.
%Instead, we propose to update the estimates
%over fixed time intervals with batch data, which can lead
%to more accurate approximation.

%It should be noted that the proposed DPSM can be regarded as the special case of the DZPSM when the zero-inflated probability $\phi_i$ is set to be 0. Thus, we only discusses the model estimation and prediction procedures for DZPSM in the following.
\subsection{Model specifications for online monitoring using batch data}
We first divide the time domain $[0,1]$ into $S$ equally spaced
intervals $[(s-1)/S, s/S]$ for $s = 1, \cdots, S$.
Let $\bm{y}_s = \left\{y_{s_m}, m = 1, 2,\cdots, n_s\right\}$ be the set of observations $y_{s_m}$ with $t_{s_m}\in[(s-1)/S, s/S]$,
and corresponding  $\bm{x}_s = \left\{x_{s_m}, m = 1,2,\cdots,n_s\right\}$.
Let $\widetilde{t}=(2s-1)/2S$ be the midpoint of the $s$th interval.
We use the batch data $(\bm{x}_s, \bm{y}_s)$ to update the model.
Specifically, we evaluate the data in the interval
$[(s-1)/S, s/S]$ at the middle point $\widetilde{t}$.
Thus, the state space model for batch data can be represented as
\begin{align}
\log \left( {{\lambda }_{s_m}} \right)&=\bm{z}_{s_m}^{T}\bm{\gamma} \left( {\tilde{t}_{s}} \right),\ \ \ \ s=1,\cdots ,S, \ m=1,2, \cdots, n_s \nonumber\\
\bm{\gamma} \left( \tilde{t}_{s} \right)&={ \tilde{\bm{T}}}\bm{\gamma} \left( \tilde{t}_{s-1} \right)+\bm{\eta} \left( {\tilde{t}_{s}},{\tilde{t}_{s-1}} \right),\ \ \bm{\eta} \left( {\tilde{t}_{s}},{\tilde{t}_{s-1}} \right)\sim N\left(\bm{0},{\tilde{\bm{Q}}} \right) ,
\label{eq-DPSM-3}
\end{align}
where
\begin{align*}
{\bm{z}_{s_m}}&={{\left( 1,0,{{x}_{s_m1}},0,{{x}_{s_m2}},\cdots ,{{x}_{s_m{{q}_{1}}}},0,{{x}_{s_m,{{q}_{1}}+1}},{{x}_{s_m,{{q}_{1}}+2}}\cdots ,{{x}_{s_m,q}} \right)}^{T}}, \\
\bm{\gamma} \left( t \right)&={{\left( {{\beta }_{0}}\left( t \right),{{{{\beta }'}}_{0}}\left( t \right),{{\beta }_{1}}\left( t \right),{{{{\beta }'}}_{1}}\left( t \right),\cdots ,{{\beta }_{{{q}_{1}}}}\left( t \right),{{{{\beta }'}}_{{{q}_{1}}}}\left( t \right),{{\alpha }_{1}},\cdots ,{{\alpha }_{{{q}_{2}}}} \right)}^{T}}, \\
\tilde{{\bm{T}}}&=\text{diag}\left( \overbrace{{\tilde{{T}}},\cdots ,{\tilde{{T}}}}^{{{q}_{1}}+1},\overbrace{1,\cdots ,1}^{{{q}_{2}}} \right),
\tilde{T}=
\begin{bmatrix}
1 & 1/S\\
0 & 1
\end{bmatrix},\\
\tilde{\bm{Q}}&=\text{diag}\left( \overbrace{\lambda _{0}^{-1}{\tilde{Q}},\cdots ,\lambda _{{{q}_{1}}}^{-1}{\tilde{Q}}}^{{{q}_{1}}+1},\overbrace{0,\cdots ,0}^{{{q}_{2}}} \right), \tilde{Q} =\begin{bmatrix}
(1/S)^3/3 & (1/S)^2/2\\
(1/S)^2/2 & 1/S
\end{bmatrix} .
\end{align*}
\section{Model estimation and prediction}\label{sec: estimation}

\subsection{Estimation and Prediction}
Due to the recursive nature of state space models, for the given smoothing parameters, the model coefficients can be efficiently estimated using a Kalman filter algorithm composed of a prediction step and a filtering step at each time point. The likelihood can also be calculated and maximized to estimate the smoothing parameters using the same algorithm. Given the estimates of the smoothing parameters, the K-step-ahead prediction can be done by running the Kalman filter prediction steps without the filtering steps.  When the new data becomes available, the coefficients can be efficiently updated by running the algorithm from the original estimates over the new observations.
{We first outline the whole picture of our algorithm which include the following steps:
	\begin{itemize}
		\item[(A)] Fit model on the training data set, \\
		(a) Select the smoothing parameters based on likelihood.\\
		(b) Using Kalman filter algorithm to estimate coefficients.
		\item[(B)] K-steps Kalman filter ahead prediction.
		\item[(C)] When new data comes:\\
		(a) Update smoothing parameters based on the likelihood of all available data.\\
		(b) Doing filtering step to estimate the coefficients on the current timepoint.
		\item[(D)] Iterate (B)-(C).
	\end{itemize}
%	\textit{Remark: In step (C), smoothing parameters could be updated when each new data is available. But because smoothing parameter controls the smoothness feature of a function which is a global feature of function and usually does not changed frequently. Thus, we recommend to update the smoothing parameters in a certain period of time. In our real data analysis, we update smoothing parameters every six months.}
		\begin{remark}
			In step (C), smoothing parameters could be updated when each new data is available. But because smoothing parameter controls the smoothness feature of a function which is a global feature of function and usually does not changed frequently. Thus, we recommend to update the smoothing parameters in a certain period of time.
			%In our real data analysis, we update smoothing parameters every six months.
		\end{remark}
}
{We next describe the Kalman filter, the maximum likelihood estimates for the smoothing parameters and the dynamic prediction steps.}
\subsection{Estimation}
In this section we describe how to fit our state-space model using a Kalman filter algorithm.
Let $\bm{Y}^{s}=({\bm{y}}_{1},\cdots,{\bm{y}}_{s})$ and $\bm{X}^s=(\bm{x}_1,\cdots,\bm{x}_s) $ be the set of past observations of the outcome and covariates up to time $t=s/S$. {
	For given smoothing parameters $\lambda_j,\ j=0,1,\cdots,q_1$, the algorithm sequentially repeats the prediction step and filtering step to  estimate the coefficients. %we can do the following estimation procedure.
	For simplicity of notation, denote $\bm{\gamma}_s\equiv\bm{\gamma}(\widetilde{t}_{s})$.
	The general algorithm of model fitting proceeds in the following steps:
	%Kalman filter is a recursive estimation allows for sequential, online processing which is done in two steps: prediction and filtering.
	\begin{enumerate}
		\item[(i)] Start with $s=0$, take diffuse prior $p(\bm{\gamma}_0|\bm{Y}^0)\sim N(\mathbf{0},c\mathbf{I})$ as the initial prior distribution of $\bm{\gamma}_1$, where $c$ is a large constant.
		\item[(ii)] Prediction step: calculate \begin{eqnarray*}
			&&\mathbb{E}(\bm{\gamma}_{s}|\bm{Y}^{s-1},\bm{X}^{s-1})=\mathbf{\widetilde{T}}\mathbb{E}(\bm{\gamma}_{s-1}|\bm{Y}^{s-1},\bm{X}^{s-1}),\nonumber\\
			&&\text{Var}(\bm{\gamma}_{s}|\bm{Y}^{s-1},\bm{X}^{s-1})=\mathbf{\widetilde{T}}\text{Var}(\bm{\gamma}_{s-1}|\bm{Y}^{s-1},\bm{X}^{s-1})\mathbf{\widetilde{T}}^\prime+\mathbf{\widetilde{Q}}.
		\end{eqnarray*}
		%$\mathbb{E}({\bm{\gamma}}_{s}|{\bm{Y}}^{s-1},{\bm{X}}^{s-1})=\mathbf{\widetilde{T}}\mathbb{E}({\bm{\gamma}}_{s-1}|{\bm{Y}}^{s-1},{\bm{X}}^{s-1}),$ and
		%$\text{Var}(\bm{\gamma}_{s}|\bm{Y}^{s-1},\bm{X}^{s-1})=\mathbf{\widetilde{T}}\text{Var}({\bm{\gamma}}_{s-1}|{\bm{Y}}^{s-1},{\bm{X}}^{s-1})\mathbf{\widetilde{T}}'
		%+\mathbf{\widetilde{T}}\mathbf{\widetilde{Q}}\mathbf{\widetilde{T}}'$.
		%$\mathbb{E}(\bm{\gamma}_{s}|\bm{Y}^{s-1},\bm{X}^{s-1})$ and $\text{Var}(\bm{\gamma}_{s}|\bm{Y}^{s-1},\bm{X}^{s-1})$.
		\item[(iii)] Filtering step: calculate the posterior mean $\mathbb{E}(\bm{\gamma}_s|\bm{Y}^{s-1},\bm{X}^{s-1},\bm{y}_{s},\bm{x}_s)$ and
		variance\\ $\text{Var}(\bm{\gamma}_s|\bm{Y}^{s-1},\bm{X}^{s-1},\bm{y}_{s},\bm{x}_s)$ through normal approximation.
		\item[(iv)] Repeat step ii) and iii)  successively for $s=1,2,\cdots,S$.
\end{enumerate}}
%when new dataset $(\bm{x}_{s},\bm{y}_{s})$ comes in.
%implements two steps: a prediction step and filtering step.
%\begin{enumerate}
%\item[(b)] Kalman filter:\\
%(i) Prediction step: calculate $\mathbb{E}(\bm{\gamma}_{s}|\bm{Y}^{s-1},\bm{X}^{s-1})$ and $\text{Var}(\bm{\gamma}_{s}|\bm{Y}^{s-1},\bm{X}^{s-1})$.\\
%(ii) Filtering step: calculate posterior mean $\mathbb{E}(\bm{\gamma}_s|\bm{Y}^{s-1},\bm{X}^{s-1},\bm{y}_{s},\bm{x}_s)$ and
% variance\\ $\text{Var}(\bm{\gamma}_s|\bm{Y}^{s-1},\bm{X}^{s-1},\bm{y}_{s},\bm{x}_s)$. %when new dataset $(\bm{x}_{s},\bm{y}_{s})$ comes in.
%\end{enumerate}

In the filtering step,
%parameters are updated through calculating posterior mean $\mathbb{E}(\bm{\gamma}_s|\bm{Y}^{s-1},\bm{y}_{s})$ and
%posterior variance $\text{Var}(\bm{\gamma}_s|\bm{Y}^{s-1},\bm{y}_{s})$ when new dataset $(\bm{x}_{s},\bm{y}_{s})$ comes in.
the posterior distribution of coefficients $\bm{\gamma}_s$ can be written as
\begin{align}\label{posterior}
p\big(\bm{\gamma}_s|\bm{Y}^{s-1},\bm{X}^{s-1},\bm{y}_{s},\bm{x}_s\big)
&\propto p\big(\bm{y}_{s}|\bm{\gamma}_s,\bm{x}_s\big)p(\bm{\gamma}_s|\bm{Y}^{s-1},
\bm{X}^{s-1}) \nonumber\\
&=\prod_{i=s_1}^{s_{n_s}}p\left(y_{i}|{\bm{\gamma}}_s,x_i\right)\cdot p\left({\bm{\gamma}}_s|{\bm{Y}}^{s-1},\bm{X}^{s-1}\right).
\end{align}
%in which $p(\bm{\gamma}_s|\bm{Y}^{s-1},\bm{X}^{s-1})$ acts as the prior and $p\big(\bm{y}_{s}|\bm{\gamma}_s,\bm{x}_s\big)$ the likelihood.
%When $s=1$, $p(\bm{\gamma}_s|\bm{Y}^{s-1})=p(\bm{\gamma}_1|\bm{Y}^{0})$ which is a prior distribution before having any data.
%According to Wahba (1978), a non-informative diffuse normal distribution with mean zero and diffuse variance is used as prior distribution.
Here $p(\bm{\gamma}_s|\bm{Y}^{s-1},\bm{X}^{s-1})$ is a normal distribution with mean and variance given by $\mathbb{E}(\bm{\gamma}_s|\bm{Y}^{s-1},\bm{X}^{s-1})$  and $\text{Var}(\bm{\gamma}_s|\bm{Y}^{s-1},\bm{X}^{s-1})$ respectively.
{However, since $p\big(\bm{y}_{s}|\bm{\gamma}_s,\bm{x}_s\big)$ is a product of poisson distribution,
	the posterior distribution (\ref{posterior}) does not have a closed-form expression.
	We approximate the posterior distribution with a normal distribution where the mean of the approximating normal distribution is the mode of posterior distribution.}
%Here we use a binomial distribution of parameter $\widetilde{\bm{\gamma}}_m$ to approximate integration of heterogeneous binary distributions.
%As we take a large $M$, this parameter approximation will be good enough.

Denote
\begin{equation}
{\varphi\left({\bm{\gamma}}_s\right)}=\sum_{i=s_1}^{s_{n_s}}\log p\left(y_{i}|{\bm{\gamma}}_s,x_i\right)+\log p\left({\bm{\gamma}}_s|{\bm{Y}}^{s-1},{\bm{X}}^{s-1}\right).
\end{equation}
The Newton-Raphson method can be used to maximize $\varphi\left({\bm{\gamma}}_s\right)$ to get its mode.
The starting value is taken as ${\bm{\gamma}}^{(0)}_s=\mathbb{E}({\bm{\gamma}}_s|{\bm{Y}}^{s-1},{\bm{X}}^{s-1})$.
%Take  as the starting value.
The procedure is to iterate the following equation until convergence with $k$-th iteration being
\begin{eqnarray}
%\label{filtering_mean}&&\mathbb{E}(\widetilde{\bm{\gamma}}_m|\widetilde{\bm{Y}}^{m-1},\widetilde{\bm{y}}_{m})
{\bm{\gamma}}^{(k)}_s={\bm{\gamma}}^{(k-1)}_s-\left\{D^2\varphi\left({\bm{\gamma}}^{(k-1)}_s\right)\right\}^{-1}D\varphi\left({\bm{\gamma}}^{(k-1)}_s\right),\nonumber
%\label{filtering_var}&&\text{Var}(\widetilde{\bm{\gamma}}_m|\widetilde{\bm{Y}}^{m-1},\widetilde{\bm{y}}_{m})
%=-\left\{D^2\ell\left(\bm{\gamma}_*\right)\right\}^{-1}
\end{eqnarray}
where $D\varphi$ and $D^2\varphi$ are first and second derivative operator. Let ${\bm{\gamma}}^{*}_s$ be the converged value, the posterior mean and variance are
\begin{eqnarray}\label{filtering}
&&\mathbb{E}(\bm{\gamma}_s|\bm{Y}^{s-1},\bm{X}^{s-1},\bm{y}_{s},\bm{x}_{s})={\bm{\gamma}}^{*}_s,\nonumber\\
&&\text{Var}(\bm{\gamma}_s|\bm{Y}^{s-1},\bm{X}^{s-1},\bm{y}_{s},\bm{x}_{s})=\{-D^2\ell({\bm{\gamma}}^{*}_s)\}^{-1}.
\end{eqnarray}
%Hence, our state space model (\ref{model}) can be estimated very quickly through recursively iterate (\ref{prediction}) and (\ref{filtering}) for $s=1,2,\cdots,S$.

%In the prediction step, for $s=1,2,\cdots,S$, it is easy to get the following equation from transition equation (\ref{state_equ}),
%Then, combining equation (\ref{filtering}) the prediction distribution is
%\begin{eqnarray}\label{prediction}
%\bm{\gamma}_{s}|\bm{Y}^{s-1},\bm{X}^{s-1}\sim N\left(\mathbf{\widetilde{T}}{\bm{\gamma}}^{*}_{s-1},\mathbf{\widetilde{T}}\{-D^2\ell({\bm{\gamma}}^{*}_{s-1})\}^{-1}\mathbf{\widetilde{T}}^\prime+\mathbf{\widetilde{Q}}\right)
%\end{eqnarray}

%\textcolor{blue}{
%Thus, the procedure of repeating (b)?C(c) successively for $s=1,2,\cdots,S$ sequentially optimize following likelihood with respect to the parameters $\bm{\gamma}_{1},\cdots,\bm{\gamma}_{s}$,
%\begin{eqnarray*}
%p(\bm{y}_{1},\cdots,\bm{y}_{S}|\bm{\gamma}_{1},\cdots,\bm{\gamma}_{S},\bm{X}^{S}) \propto
%\Pi_{s=1}^{S} \left\{p(\bm{y}_{s}|\bm{x}_{s},\bm{\gamma}_{s})p(\bm{\gamma}_{s}|\bm{Y}^{s-1},\bm{X}^{s-1})\right\},
%\end{eqnarray*}
%where $p(\bm{\gamma}_{1}|\bm{Y}^{0})=p(\bm{\gamma}_{1})$ is initial prior as in step (a). Note that $\bm{\gamma}_{s}$ only exists in $s$-th product item $p(\bm{y}_{s}|\bm{x}_{s},\bm{\gamma}_{s})p(\bm{\gamma}_{s}|\bm{Y}^{s-1},\bm{X}^{s-1})$ which makes the estimation a sequential procedure. }

\subsection{Smoothing parameter selection}\label{subsec: smoothing}
In the model,  the smoothing parameters $\lambda_j$ control the smoothness of $\beta_j(t),\ j=0,1,\cdots,q_1$ respectively.
The selection of $\lambda_j$ plays a key role in model fitting and prediction.
We propose to maximize the following log-likelihood to estimate the smoothing parameters,
\begin{eqnarray}
(\widehat{\lambda}_0,\cdots,\widehat{\lambda}_{q_1})=\arg \max_{{\lambda}_0,\cdots,{\lambda}_{q_1}}\sum_{s=1}^{S}
\sum_{i=s_1}^{s_{n_s}}\log p(y_{i}|{\bm{Y}}^{s-1},{\bm{X}}^{s-1}),\nonumber
\end{eqnarray}
where
\begin{eqnarray}
p(y_{i}|{\bm{Y}}^{s-1},{\bm{X}}^{s-1})=\int p(y_{i}|{\bm{\gamma}}_{s},{\bm{Y}}^{s-1},{\bm{X}}^{s-1})p({\bm{\gamma}}_{s}|{\bm{Y}}^{s-1},{\bm{X}}^{s-1})d{\bm{\gamma}}_{s},\nonumber
\end{eqnarray}
and is not available in closed form. %Laplace approximation is commonly used tool to circumvent complicated integral (Lewis and Raftery, 1997).
A commonly used method is Laplace approximation.
\cite{lewis1997estimating} suggest that this approximation should be quite accurate.
The Laplace approximation yields
\begin{eqnarray}
p(y_{i}|{\bm{Y}}^{s-1},{\bm{X}}^{s-1})\approx (2\pi)^{(q+q_1+2)/2}\left|\{-D^2\ell_i({\bm{\gamma}_{s}^*})\}^{-1}\right|^{1/2}p(y_{i}|{\bm{\gamma}}^*_{s}, {\bm{Y}}^{s-1},{\bm{X}}^{s-1})p({\bm{\gamma}}_{s}^*|{\bm{Y}}^{s-1},{\bm{X}}^{s-1})\nonumber
\end{eqnarray}
where $\ell_i(\bm{\gamma})=\log p(y_{i}|{\bm{\gamma}},{\bm{Y}}^{s-1},{\bm{X}}^{s-1})+\log p\left({\bm{\gamma}}|{\bm{Y}}^{s-1},{\bm{X}}^{s-1}\right)$.
%We can then directly optimize the function using existing packages in R, such as \textit{optim} function.
%The smoothing parameters can be estimated using existing data and fixed in the prediction. When new  data becomes available, the likelihood needs to be updated to estimate the smoothing parameters.
	%, which can be efficiently obtained by running the proposed algorithm over the new observations only.
Although the smoothing parameters can be updated whenever new data becomes available, we typically do not update them frequently in practice because the curve smoothness often does not change much over time.
%Although the smoothing parameters can be updated as new data are available, we found it is not necessary to update it very frequently.
%A long training period usually captures the curvature of the functional curve well, which typically does not change for a long period of time.  So we recommend that the estimates $\widehat{\lambda}_0,\cdots,\widehat{\lambda}_{p_1}$ are used in future predictions \overlinewithout further updating.

\subsection{K-step-ahead Prediction}%{Dynamic Prediction}%\label{observational_equ}

The K-step-ahead predicted model coefficients from timepoint $\widetilde{t}_{S}$ follow
%\begin{eqnarray}
%\mathbb{E}({\bm{\gamma}}_s|{\bm{Y}}^{s-1})={T}\mathbb{E}({\bm{\gamma}}_{s-1}|{\bm{Y}}^{s-1}),\nonumber\\
%\text{Var}({\bm{\gamma}}_s|{\bm{Y}}^{s-1})={T}\text{Var}({\bm{\gamma}}_{s-1}|{\bm{Y}}^{s-1}){T}^\prime+\widetilde{Q},\nonumber
%\end{eqnarray}
\begin{eqnarray}
&&\mathbb{E}({\bm{\gamma}}_{S+K}|{\bm{Y}}^{S},{\bm{X}}^{S})=\mathbf{\widetilde{T}}^KE({\bm{\gamma}}_{S}|{\bm{Y}}^{S},{\bm{X}}^{S}),\nonumber\\
&&\text{Var}({\bm{\gamma}}_{S+K}|{\bm{Y}}^{S},{\bm{X}}^{S})=\mathbf{\widetilde{T}}^K \text{Var}({\bm{\gamma}}_{S}|{\bm{Y}}^{S},{\bm{X}}^{S}){{\mathbf{\widetilde{T}}^{\prime K}}}+\mathbf{\widetilde{T}}^{K-1}{\mathbf{\widetilde{Q}}}\mathbf{\widetilde{T}}^{\prime K-1}+\cdots+\mathbf{\widetilde{T}}\mathbf{\widetilde{Q}}{\mathbf{\widetilde{T}}^\prime}+\mathbf{\widetilde{Q}}.\nonumber
\end{eqnarray}
{The predicted intensity with the given covariates can then be calculated using  the coefficients with the predicted expectations.}
% $\mathbf{z}$ at time $\widetilde{t}_{S+K}$ is
%\begin{equation}
%\color{blue}{p(y=1|\mathbf{z})=\frac{\exp\{\mathbf{z}^TE({\bm{\gamma}}_{S+K}|{\bm{Y}}^{S})\}}{1+\exp\{\mathbf{z}^TE({\bm{\gamma}}_{S+K}|{\bm{Y}}^{S})\}}}
%\end{equation}
%where $z$ is produced by covariates $x$ through the same way with $\mathbf{z}_{s_m}$ in model (\ref{observational_equ}).
%When data at time $\widetilde{t}_{S+1}$ are available, we can estimate
% $\bm{\gamma}_{S+1}$ using the same Kalman filter algorithm in model estimation part.
%through calculating  $\mathbb{E}({\bm{\gamma}}_s|{\bm{Y}}^{s-1},\bm{y}_{s})$ and  $\text{\text{Var}}({\bm{\gamma}}_s|{\bm{Y}}^{s-1},\bm{y}_{s})$ by (\ref{filtering}).
%With cubic smoothing spline, K-step ahead prediction of parameters is linear extrapolation. This linear extrapolation includes the trend of information in historical data, which enables the new proposed model to have higher prediction accuracy.

\section{Simulation}\label{sec: simulation}
In this section, we evaluate the predictive performance of the proposed model using simulated data.
We compare our proposed method with generalized linear models in which all coefficients are constant (referred to as the Poisson, NB and ZIP method correspond to Poisson, negative-binomial and zero-inflated Poisson).

We simulate a count outcome using the following model,
\begin{eqnarray}
P(Y_i=k)=\frac{\lambda_i^ke^{-\lambda_i}}{k!},\ \mathrm{log}(\lambda_i)=\beta_0(t_i)+\beta_1(t_i)X_{i1}+\alpha X_{i2},  \nonumber
\end{eqnarray}
where $t_i$ is generated from a uniform distribution $U[0,1]$ and $X_{i1}$ and $X_{i2}$ are independently generated from a uniform distribution $U[0,1]$.
$\beta_0(t)$  is the varying intercept generated by $\beta_0(t)=t-2,$ and $\beta_1(t)$ is the varying coefficient for $X_{i1}$ generated by $\beta_1(t)=0.2\log(t)+0.5$ , $\alpha=0.25$ is a constant coefficient.
The average intensity (expected claim frequency), i.e., $\frac{1}{n}\sum_{i=1}^{n}\lambda_i$,  over time is around 0.24.
We evaluate the model performance under sample size $n=10^5$.
The first 75\% of the data is used to fit the model, and the remaining 25\% is used for testing model performance.

\begin{figure}[h!]
	\begin{center}
		\includegraphics[width=17cm,height=4.5cm]{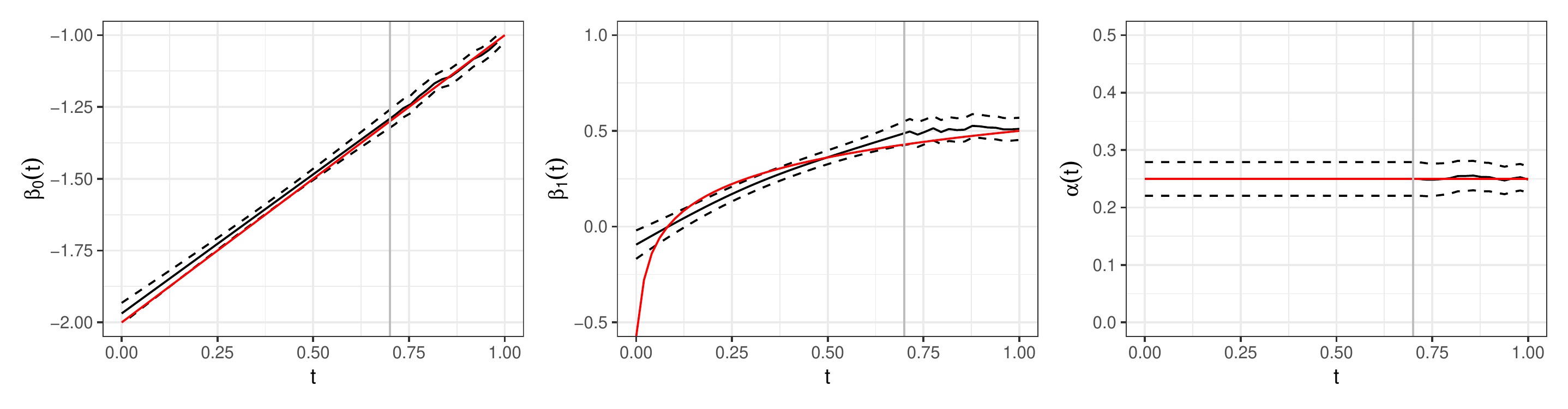}
	\end{center}
	\caption{
		The plot of coefficients for the intercept (\textit{left panel}), $X_{1}$ (\textit{middle panel}), and $X_2$ (\textit{right panel}) with 95\% confidence bands using the proposed method. The sample size is  $n=10^5$. The red line represents the true value for each parameter. The coefficient plots are divided at $t= 0.75$, corresponding to the estimated coefficient in the training data and one-step ahead prediction in the validation data, respectively.
	}
\label{fig-simulation-coefficients}
\end{figure}

For our proposed DPSS model, we first divide the data into batches. In the data set, we divide the time into $50$ equally spaced time intervals, $[(s-1)/50,s/50], s=1,2, \cdots , 50$. On average, there are $2000$ subjects within each time interval, corresponding to the total sample size of $10^5$.
We use a normal distribution $N(0,100\cdot I_5)$ as the initial prior distribution of the coefficients in the model,
where $I_5$ is the 5-dimensional identity matrix.
{We use the proposed criterion as in section \ref{subsec: smoothing} to select the smoothing parameters for $\beta_0(t)$ and $\beta_1(t)$ using the first 75\% data, denoted as $\lambda_0$ and $\lambda_1$ respectively.}
Both $\lambda_0$ and $\lambda_1$ are fixed when making predictions in the remaining data.

We compare the estimated functional curves and corresponding 95\% confidence bands for each coefficient of our method. Figure \ref{fig-simulation-coefficients} shows the estimated coefficients for the intercept (\textit{left panel}), $X_1$ (\textit{middle panel}), and $X_2$ (\textit{right panel}) with corresponding 95\% confidence  bands. The coefficient plots are divided at  $t=0.75$ indicated by vertical dotted line, corresponding to the training and validation data, respectively. The figure shows the estimated coefficients in the training set ($t\leq0.75$) and one-step ahead predicted coefficients in the validation set ($t>0.75$). The estimated coefficient curves using our proposed method are close to the true curves (red lines). In addition, the 95\% confidence bands cover the true values in both the training and validation data.

We evaluate the accuracy of the predicted claim of outcome using two different metrics. The first one is Poisson Deviance, which measures how closely the fitted model¡¯s predictions are to the observed outcomes. It is defined as
\begin{equation}
\label{eq-poisson-deviance}
D(y_i, \hat{y}_i)=2\sum_{i=1}^n\left[y_i\log(y_i/\hat{y}_i)-(y_i-\hat{y}_i) \right],
\end{equation}
where $\hat{y}_i$ denotes the predicted mean for on observation $i$ based on the estimated model parameters. A smaller Poisson Deviance means a better model fit.
The second one is the difference between predicted and observed data, which is defined as
\begin{equation*}
D(\hat{\lambda}_i, y_i)=\sum_{i=1}^{n_{test}}I(y_i=k)-\sum_{i=1}^{n_{test}}\frac{\hat{\lambda}_i^k e^{-\hat{\lambda}_i}}{k!},\ k=0,1,2,\cdots,
\end{equation*}
where $n_{test}$ is the sample size of test data and $\hat{\lambda}_i$ corresponding predicted claim frequency.
Table \ref{tab-sim-dev} and Table \ref{tab-simulation-predicted} show that our proposed method has much better prediction performance compared with static models (Poisson, NB and ZIP) as
the static models are not able to capture the varying effects over time and do not perform well, as one would expect.

%
%We also evaluate the accuracy of the predicted claim of outcome using following difference criterion.
%The first one is Poisson Deviance, which measures how closely the fitted model¡¯s predictions are to the observed outcomes. It is defined as
%$$
%D(y_i, \hat{y}_i)=2\sum_{i=1}^n\left[y_i\log(y_i/\hat{y}_i)-(y_i-\hat{y}_i) \right],
%$$
%where $\hat{y}_i$ denotes the predicted mean for on observation $i$ based on the estimated model parameters. A smaller Poisson Deviance means a better model fit.

%Table \ref{tab-simulation-deviance} shows the Poisson Deviance for four competing models.

%The second one is the difference between predicted and observed data in

\begin{table}
	\caption{The Poisson deviance for the testing data in the simulated data set.}
	\label{tab-sim-dev}
	\begin{center}
		%		\begin{tabular*}{\hsize}{@{}@{\extracolsep{\fill}}c|c|c|c|c @{}}
\begin{tabular}{ccccc}
	\toprule
Models	&    Poisson &         NB &        ZIP &       DPSS \\
	\hline
Deviance &     0.9354 &     0.9354 &    0.9178 &     0.8557 \\
	\bottomrule
\end{tabular}

		%		\end{tabular*}
	\end{center}
\end{table}

\begin{table}[htbp]
	\caption{The difference between predicted and observed claim frequency for the testing data in the simulated data set.}\label{tab-simulation-predicted}
%	\begin{tabular*}{\hsize}{@{}@{\extracolsep{\fill}}c|c|c|c|c @{}}
\centering
	\begin{tabular}{c|c|c|c|c }
		\toprule
		Count $(k)$&Poisson & NB & ZIP &DPSS \\
		\hline
         0 &       3015 &       3055 &       3055 &         24 \\
1 &      -2139 &      -2209 &      -2213 &        -58 \\
2 &       -750 &       -727 &       -720 &          9 \\
3 &       -112 &       -105 &       -107 &         21 \\
4 &        -14 &        -13 &        -13 &          3 \\
6 &         -1 &         -1 &         -1 &         -1 \\
		\bottomrule
			\end{tabular}
%	\end{tabular*}
\end{table}

\section{Empirical analysis}\label{sec: applications}
The performance of the DPSS model is tested by using a novel automobile insurance data set in China, which is collected over the years from 2010 to 2015.
This data set includes self-reported risk characteristics and claim history with the number of claims, of around 440,000 insurance contracts.
The claim data set is collected from three types of automobile insurance, including collision insurance, third-party liability insurance, and compulsory traffic insurance.
Table \ref{tab-claim-frequency} presents the distribution of the number of claims over time.
We observe that most
policyholders do not have a claim (82.10\%), some have one or two claims (16.4\%)
and the remaining policyholders have more than three claims.
The probability of having no claim is increasing over years, especially from 2014 to 2015.
We argue that the Chinese regulator has launched a comprehensive 2nd reform of the automobile insurance segment from 2007 to 2015,
followed by a 3rd reform from 2015 to 2020, covering pricing, fee structures, and product coverage among others, which had a big impact on the distribution of the number of claims.
Especially around 2014-2015, China launched the auto insurance reform by expanding insurance coverage,
adopting a uniform ratemaking mechanism to address the problem of  ``high premiums and low compensation" in the Chinese automobile insurance market.
This changing regulatory policy motivates us to take the time trend into the statistical models when we aim to predict the claim frequency beyond the sample period.

\begin{figure}[h!]
	\begin{center}
		\includegraphics[scale = 0.4]{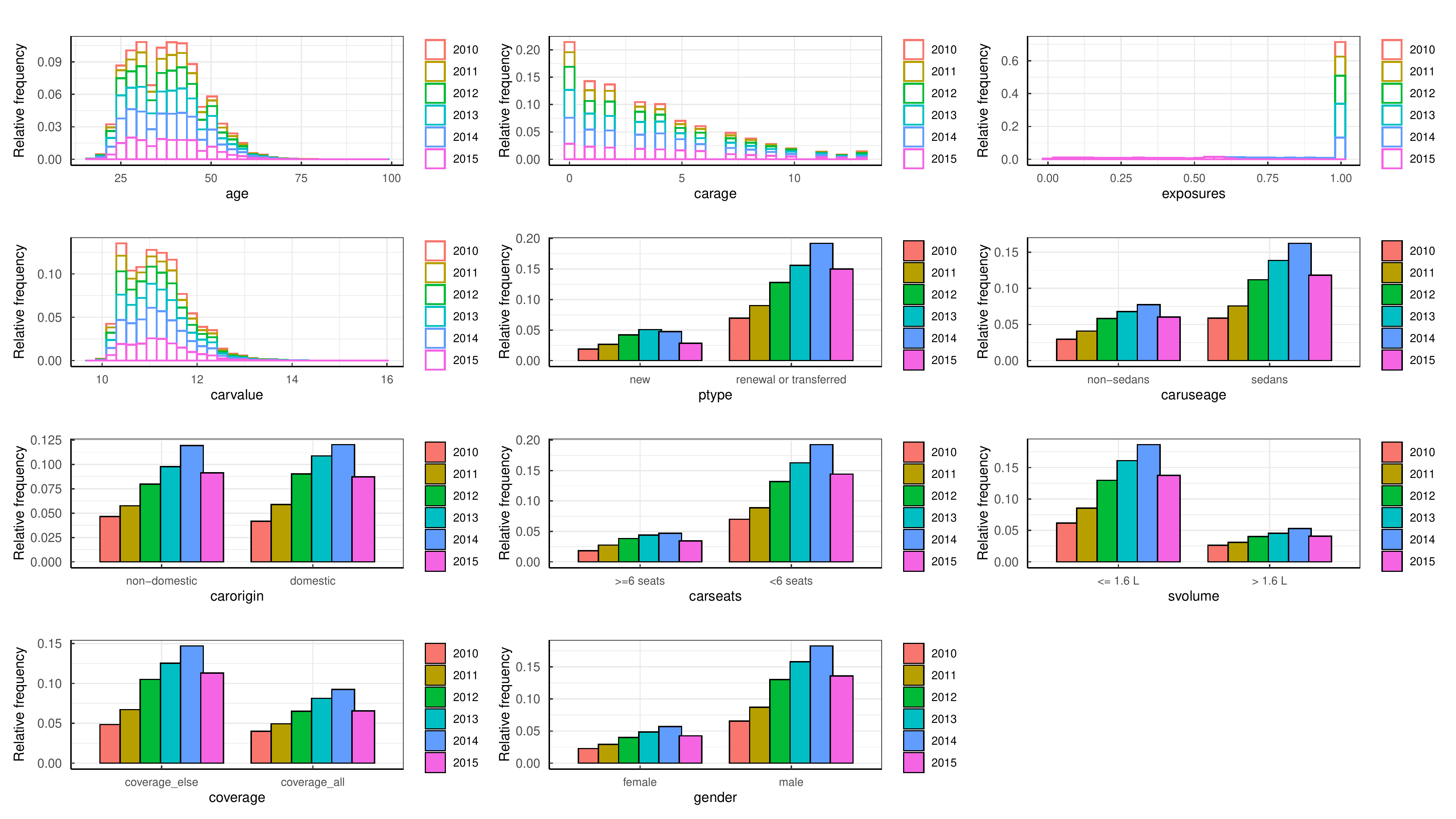}
	\end{center}
	\caption{
		Relative frequency of the covariates \textit{expo}, \textit{gender}, \textit{age}, \textit{carage},
		\textit{carvalue}, \textit{ptype}, \textit{carusage}, \textit{carorigin}, \textit{svolume}, \textit{carseats} and \textit{coverage} over years.}
		\label{fig-relative-frequency-covariates}
\end{figure}

The information on risk characteristics of the individual policies in this data set contains:
\textit{expo}, \textit{gender}, \textit{age}, \textit{carage},
\textit{carvalue}, \textit{ptype}, \textit{carusage}, \textit{svolume}, \textit{carseats} and \textit{coverage}.
The summary statistics of these variables are shown in Table \ref{tab-claim-factors}.
Figure \ref{fig-relative-frequency-covariates} shows how the covariates are distributed in the automobile insurance data set over years.
Most policyholders (70.90\%) are exposed to the risk of filing a claim during a whole policy year, while the others
between zero and one, which indicates that the underwriting date of those policies might be after the start of the year, or surrender of the contract occurs before the end of the year. Almost
all policyholders (89.12\%) are aged between 25 and 60, which means that there are few young and
old policyholders in the insurance portfolio, and more than half of the policyholders (75.96\%) are female.
Most policies (78.59\%) are renewed or transferred from other insurance companies.
More than half of vehicles (76.20\%) have a displacement of less than 1.6 L,
half of the vehicles (50.73\%) are domestic,
79.05\% of vehicles have a seating capacity of fewer than 6 seats and 66.57\% of vehicles are sedans.
More than half of policies (60.60\%) do not cover all three types of insurance: collision insurance, third-party liability insurance, and compulsory traffic insurance.

\begin{table}[htbp]
	\caption{Distribution of numbers of claims over years in automobile insurance data set.}	
	\label{tab-claim-frequency}
	\begin{tabular*}{\hsize}{@{}@{\extracolsep{\fill}}cccccccc@{}}
		\toprule
	\multirow{2}{*}{Count}		& \multicolumn{6}{c}{Percentage by Year} & \\
	\cline{2-8}
	& 2010 & 2011& 2012 & 2013 &2014 &2015& Overall \\
		\hline
        0 & 77.20\% & 77.90\% & 78.70\% & 78.70\% & 82.10\% & 94.70\% & 82.10\%  \\
1 & 15.00\% & 14.80\% & 14.20\% & 14.50\% & 13.60\% & 4.70\% & 12.60\%  \\
2 & 5.10\% & 5.00\% & 5.00\% & 5.00\% & 3.50\% & 0.50\% & 3.80\%  \\
3 & 1.70\% & 1.60\% & 1.50\% & 1.40\% & 0.70\% & 0.10\% & 1.10\%  \\
4+ & 1.00\% & 0.70\% & 0.50\% & 0.40\% & 0.20\% & 0.00\% & 0.40\%  \\
\hline
obs. & 39,103 & 51,604 & 75,298 & 91,418 & 106,076 & 79,012 & 442,511  \\
		\bottomrule
	\end{tabular*}
	%	}
\end{table}

\begin{table}[htbp]
	\small
	\centering
	\caption{Description of the risk characteristics in automobile insurance data set.}
	\label{tab-claim-factors}
	\renewcommand\arraystretch{1.5}
	\begin{tabular*}{\hsize}{@{}@{\extracolsep{\fill}}lll@{}}
		\toprule
		\textbf{Variables} & \textbf{Type} & \textbf{Description}\\
		\hline
		expo & continuous & risk exposures measured by the effective policy duration: 0-1\\
		age&	continuous&	age of the policyholders: 18-99\\
		carage &	continuous & age of vehicle: 0-13\\
		carvalue & continuous & log transformation of purchase price of the vehicle: 9.8 (minimum) -16 (maximum)\\
		ptype &categorical &\tabincell{l}{ policy type (two levels):
			newly underwritten policy with new vehicle (\textit{new}), \\
		renewal policy or transferred policy from other insurance company (\textit{renewal or transferred})}
%		(\textit{renewal}), \\
%		Transferred policy from other insurance company (\textit{transferred})}
	\\
		carusage &categorical&categories of vehicle (two levels):
		\textit{sedans} and \textit{non-sedans}\\
		carorigin&categorical & two levels: \textit{domestic} or \textit{non-domestic} vehicles
		\\
		svolume &categorical &swept volume of vehicle (two levels): $>1.6L$ or $\le 1.6L$\\
		carseats &categorical &number of seats in vehicle (two levels): $\ge 6$ seats or $<6$ seats\\
		coverage &categorical & \tabincell{l}{ insurance coverage of vehicle (two levels): \\
			covering collision insurance, third-party liability insurance \\
			and compulsory traffic insurance (\textit{coverage\_all}), or
			not covering \\three types of insurance (\textit{coverage\_else})
	}\\
gender & categorical & \textit{male} or \textit{female}\\
		\bottomrule
	\end{tabular*}
\end{table}

We first fit a DPSS to model the number of claims that allows the intercept and coefficients for all the above-mentioned predictors to change over time.
Specifically, the data set is split into six batches by year.
The first five batches  (data from years 2010-2014, denoted by $\mathcal{L}(Y_{i}(k),\bm{x}_i(k))_{0\le i\le n}, k=2010,\dots,2014$)
are used as the training data for model fitting and smoothing parameter selection,
and the sixth batch data (data from 2015, denoted by $\mathcal{L}(Y_{i}(k),\bm{x}_i(k))_{0\le i\le n}, k=2015$)  is used as the testing data for an out-of-sample
analysis.
The preliminary analysis shows that only the coefficients of \textit{age}, \textit{carage} and \textit{carvalue} are constant over time
while the others keep changing over time.
Now, we consider the following DPSS:
\begin{align*}
\log \lambda_i(t_i) = &\beta_0(t_i)+\beta_1(t_i)\log \mathrm{expo}
+\beta_2(t_i) \mathrm{caruseage}_{i}+
\beta_3(t_i) \mathrm{carseats}_{i}+\beta_4(t_i) \mathrm{ptype}_{i}\\
&\quad +\beta_5(t_i) \mathrm{coverage}_{i}
+\beta_6(t_i) \mathrm{gender}_{i}+\beta_7(t_i) \mathrm{svolume}_{i}+\beta_8(t_i) \mathrm{carorigin}_{i}\\
&\quad +\alpha_1\mathrm{age}_{i}+
\alpha_2\mathrm{carage}_{i}+\alpha_3\mathrm{carvalue}_{i}.
\end{align*}
Note that binary coding is employed for the categorical variables (\textit{gender},  \textit{ptype}, \textit{carusage}, \textit{carorigin}, \textit{svolume}, \textit{carseats} and \textit{coverage}).
To illustrate the superiority of the proposed models, we also consider five competing regression models:
Poisson regression model (Poisson),
negative-binomial regression model (NB), and
zero-inflated Poisson regression model (ZIP), as well as the Dynamic negative-binomial state space model (DNBSS) and Dynamic zero-inflated Poisson state space model (DZIPSS).
The year $t_i$ as a continuous covariate is introduced into the mean parameter to capture the time trend over years in static models.
The mean of the Poisson component and the probability of extra zeros are both assumed to be modelled as the function of the covariates in the ZIP model.

\begin{figure}[h!]
	\begin{center}
		\includegraphics[width=17cm,height=14cm]{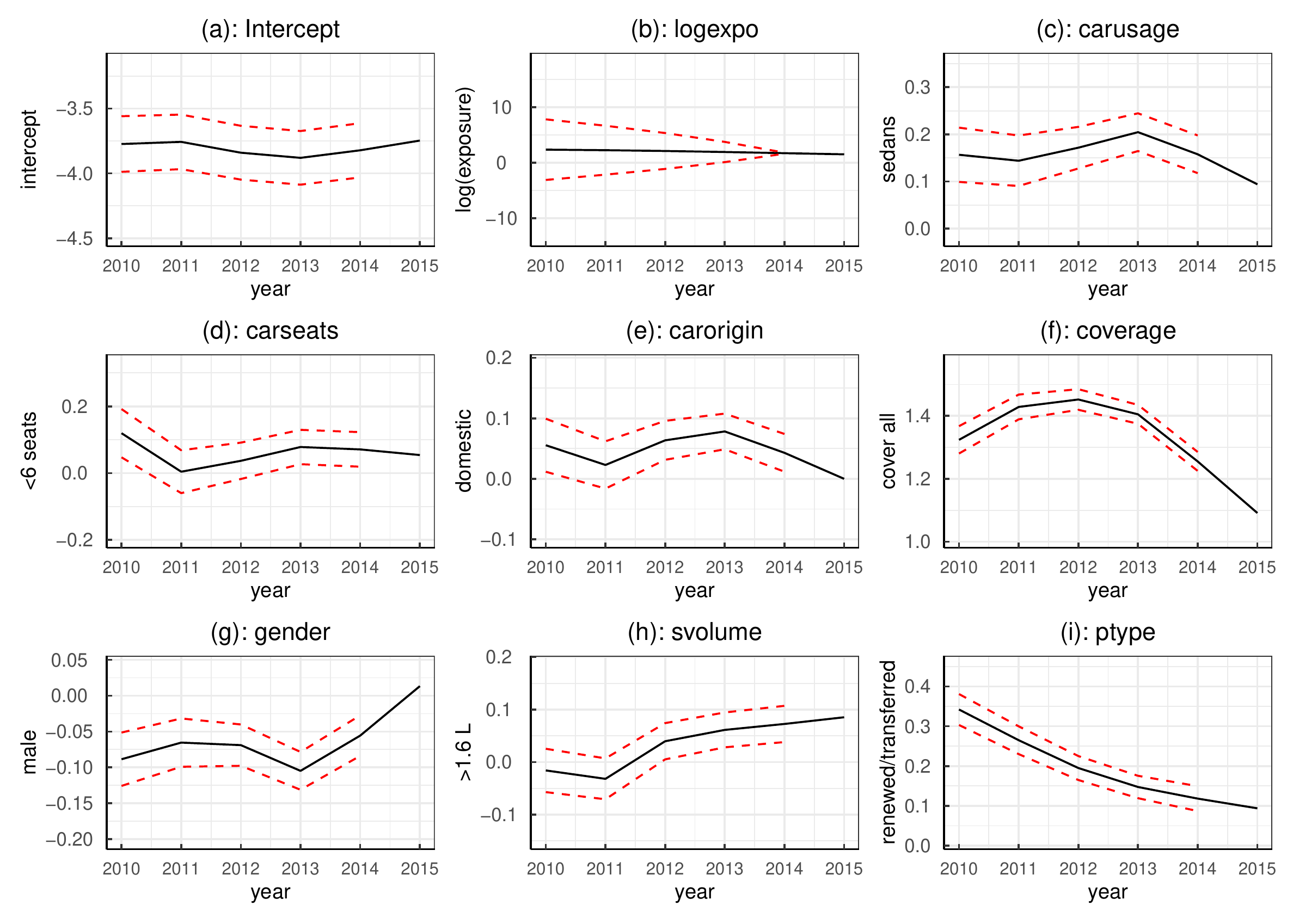}
	\end{center}
	\caption{
		The plot of coefficients for the intercept (left) and eight covariates with 95\% confidence bands using the proposed method.
		The coefficient plots are divided at $t= 0.75$, corresponding to the estimated coefficient in the training data and one-step ahead prediction in the testing data, respectively.
		\textit{Female} in \textit{gender} variable, \textit{new policy} in \textit{ptype}, \textit{non-sedans} in \textit{caruseage}, \textit{non-domestic} in \textit{carorigin}, $\le 1.6$L in \textit{svolume}, $\ge 6$ seats in \textit{carseats} and \textit{coverage\_else} in \textit{coverage}
		are used for the reference levels.
	}
\label{fig-plot-coefficients}
\end{figure}

Figure \ref{fig-plot-coefficients}(a)-(i) plot the estimated varying coefficients for the intercept and eight covariates of the DPSS model with 95\% confidence bands
as the function of years 2010-2014 in the training data and year 2015 in the testing data.
From Figure \ref{fig-plot-coefficients}(c)-(e),
we observed that the claim frequency of sedans is higher than non-sedans;
vehicles with fewer than six seats are claimed more frequently than vehicles with more than six seats;
domestic vehicles have more claims than non-domestic vehicles.
The difference in claim frequency between sedans and non-sedans has a slightly decreasing trend from 2010 to 2011, then increases from 2011 to 2013 and decreases after 2013. 
A similar time trend can be observed in Figure \ref{fig-plot-coefficients}(d) and (e).
From the time plot of the \textit{coverage} variable in Figure \ref{fig-plot-coefficients}(f), we can see that
the difference in claim frequency between policies with high insurance coverage and policies with low coverage shows an increasing trend before 2012 followed by a decreasing trend.
The positive estimated coefficient indicates that the policies with high insurance coverage have a relatively high claim frequency.
There are two possible reasons for this: (1) drivers with bad driving habits are the ones who buy insurance with high insurance coverage;
(2) the drivers with high insurance coverage drive more casually, resulting in more insurance claims.
Figure \ref{fig-plot-coefficients}(g) shows that while the female has more claims than the male, the difference in claim frequency is decreasing over time and even tends to be zero after the year 2014.
In Figure \ref{fig-plot-coefficients}(i), the decreasing time trend of \textit{ptype} can be observed, which indicates that the renewed or transferred policies have more claim frequency than new policies.
Another interesting finding is that the sign of the estimated coefficient of \textit{svolume} in Figure  \ref{fig-plot-coefficients}(h) changes from negative to positive as the year increases.
Overall, we can conduct that the time trend of most of the rating factors changes around 2011-2012,
and the impact on claim frequency is decreasing over time.
The time plots indicate that rating factors play less important roles in classification rate-making as the year increases, which is in line with the timeline of the second round of insurance rate reform that was conducted in 2007-2015 with the aim of reducing the insurance premium rate and increasing the insurance coverage.
The most important regulations for this round of reform were
the ``Motor Vehicle Commercial Insurance Model Clauses" issued by The Insurance Association of China in 2012 \footnote{\url{http://www.iachina.cn/art/2012/3/14/art_22_9744.html}},
and the ``Opinions on deepening the reform of commercial auto insurance terms and rates management system" issued by China Banking and Insurance Regulatory Commission in 2015 \footnote{\url{http://www.gov.cn/gongbao/content/2015/content_2868890.htm}}.
The uniform vehicle insurance terms and conditions were suggested by regulatory authorities during this period for automobile insurance rate-making to avoid vicious competition by insurance companies to attract customers by reducing premiums.
These changing regulations weakened the right of insurance companies to set premiums based on risk classification, which reflected a decreasing trend of estimated regression coefficients for most of the rating factors.

\begin{figure}[h!]
	\begin{center}
		\includegraphics[width=8cm,height=5cm]{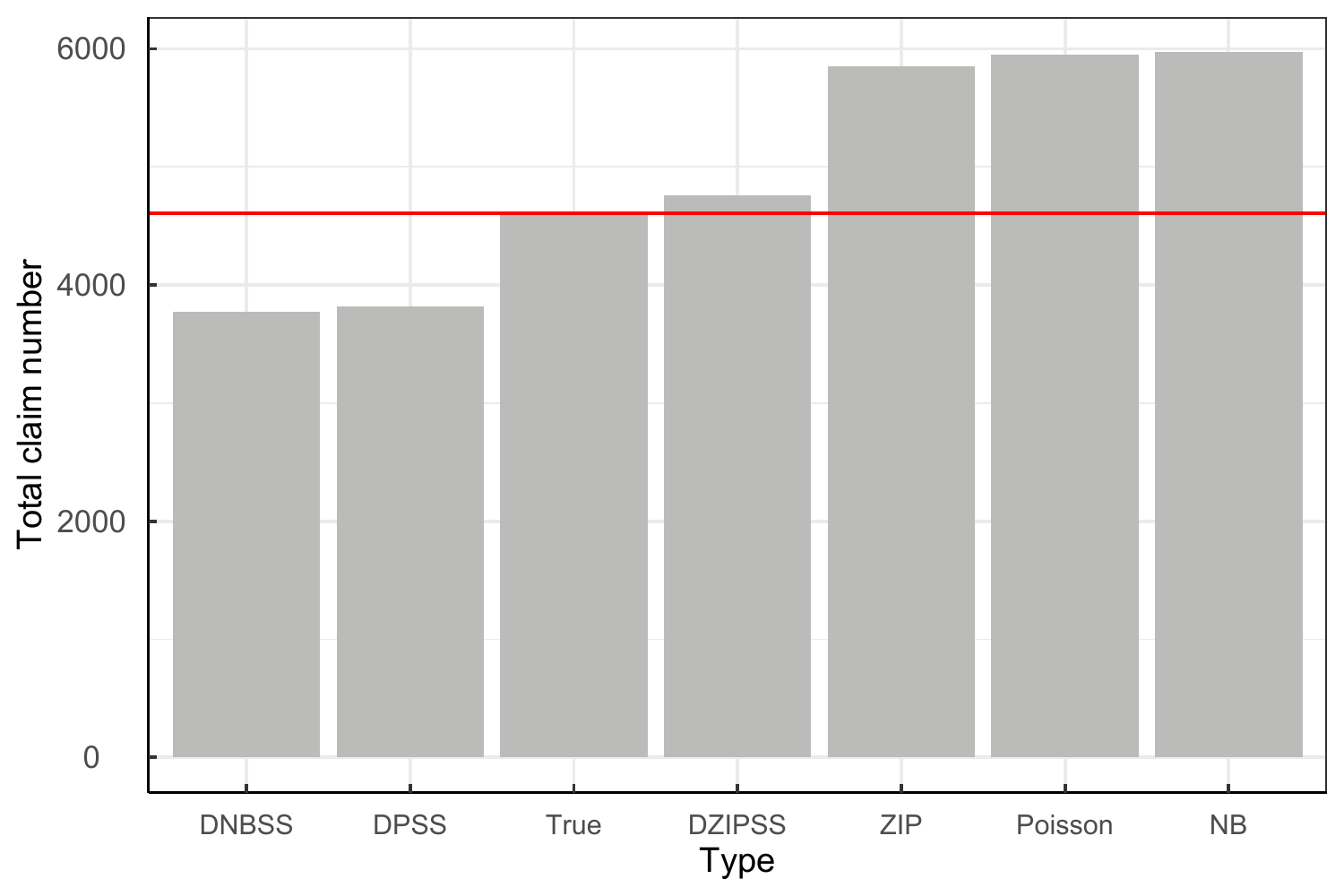}
	\end{center}
	\caption{The predicted value of the total numbers of claim for the testing data. The true observation of the total numbers of claim is 4,068, which is marked as the red line.
	}
	\label{fig-total-claim}
\end{figure}

To compare the performance of Poisson, NB, ZIP, DPSS, DNBSS, and DZIPSS models, we predict the claim frequency by applying each model on the testing data.
Table \ref{tab-diff-prediction} reports the difference between predicted and observed claims for the testing data.
One can see that our proposed dynamic prediction models have much better performance in predicting the claim frequency having no claims.
Figure \ref{fig-total-claim} shows the predicted total number of claims for the whole insurance portfolio in the testing data. 
The static prediction models (Poisson, NB, ZIP) overestimate the total number of claims, while DPSS and DNBSS underestimate the true value. DZIPSS has the best prediction performance as it has the lowest deviation.
The predicting outcomes of dynamic models are much closer to observed total number of claims than static models.
We also evaluate the accuracy of the predicted claim of outcome using the Poisson Deviance defined in \eqref{eq-poisson-deviance}, the lifts discussed in \citep{lee2021addressing}
and Gini index proposed by \citet{frees2011summarizing}.
The lifts and the Gini index are auxiliary metrics that evaluate the models without an assumption on the underlying distribution.
The lifts are the ratio of the
the average response of the top decile prediction over the average response of
the bottom decile (for two-way) and the
population (for one-way).
The lift can be interpreted as a measure of the ability of a model to differentiate observations. A higher lift illustrates that the model is more capable of
separating the extreme values from the average.
A similar auxiliary metric is the Gini index, which is a measure of assessing the discriminatory power of the models, especially
for the claim data with the high proportions of zeros and the highly right-skewed features.
%The Gini index is more robust against the fluctuation of predictive performance in both tails \citep{lee2021addressing}.
A score with a greater Gini index produces a greater separation among the observations.
In other words, a higher Gini index indicates a greater ability to distinguish good risks from bad risks.
From Table \ref{tab-dev-gini}, one can see that
dynamic models (DPSS, DNBSS, and ZIPSS) have the smaller Deviance than the static models (Poisson, NB, and ZIP), which indicates that the model fitting performance can be improved by introducing the time trend of the rating factors.
The one-way lifts for DPSS and Poisson models are 3.522 and 3.550 respectively, which indicates that
the top 10\% risk classified by the DPSS model contributes around 35.22\% of
the total claims according to the one-way lift,
while the top 10\% risk classified by static Poisson contributes around 35.50\%.
This means insurers can remove the top 10\% of risks through underwriting,
and the remaining 90\% of the policy represents only 64.78\% of the original loss according to the results of DPSS.
The insurers will charge 28.02\% (1-64.78/90) lower on average based on the DPSS model while
28.33\% (1-64.50/90)  based on the static Poisson model.
Smaller one-way and two-way lifts in dynamic models are observed, which indicates that dynamic models do not show a better ability to distinguish good risks from bad risks. The results are in line with the objectives of the second round of insurance rate reform in China that was conducted in 2007-2015 with the aim of weakening the risk differentiation ability of the rating factors.
Table \ref{tab-dev-gini} shows that there is no significant difference in the risk differentiation ability of these six models
in terms of the Gini index.

\begin{figure}[h!]
	\begin{center}
		\includegraphics[width=12cm,height=8cm]{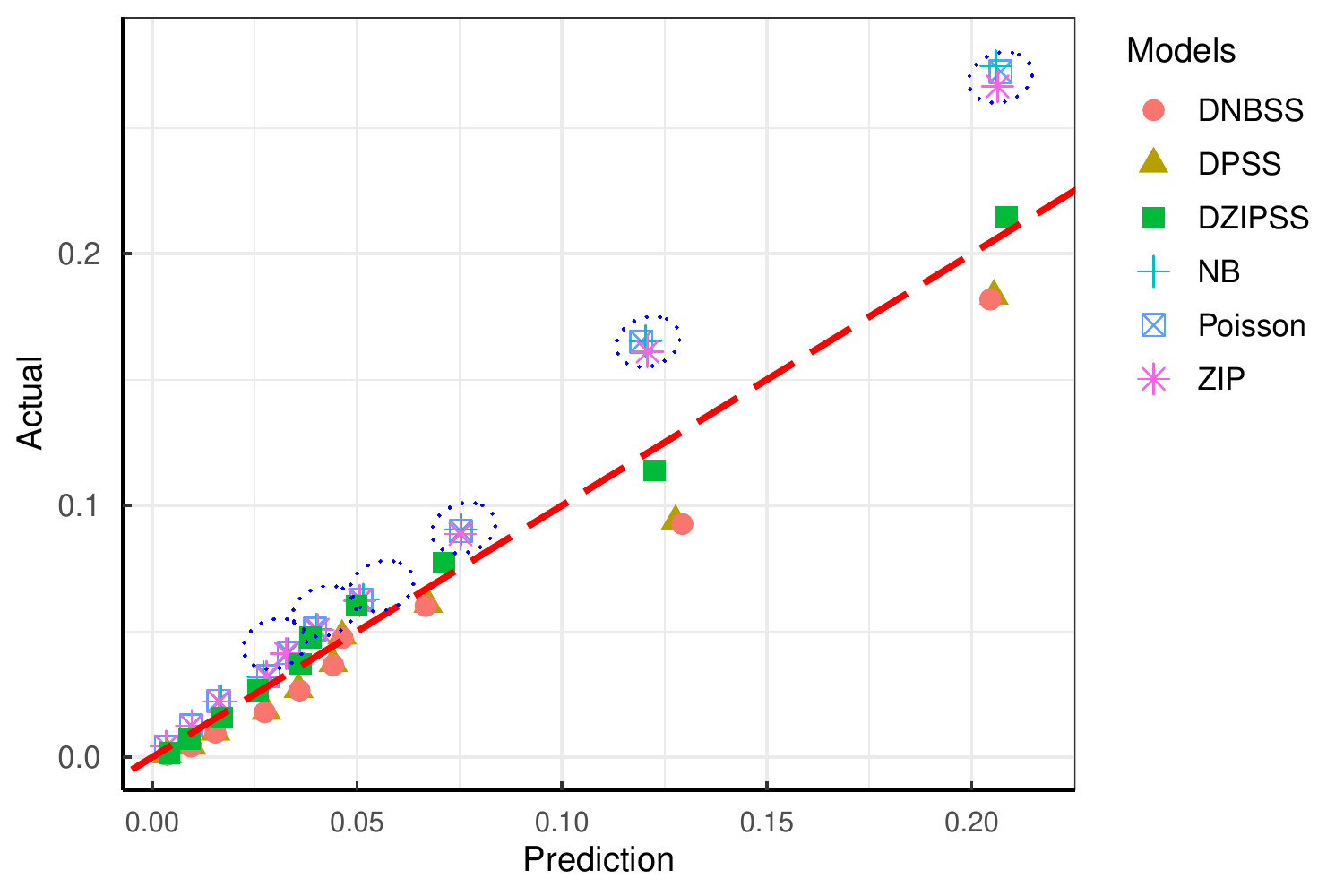}
	\end{center}
	\caption{Lift plot for competing models. Static models are circled by dark blue dotted lines.
	}
	\label{fig-lift-plot}
\end{figure}

\begin{figure}[h!]
	\begin{center}
		\includegraphics[width=12cm,height=6.5cm]{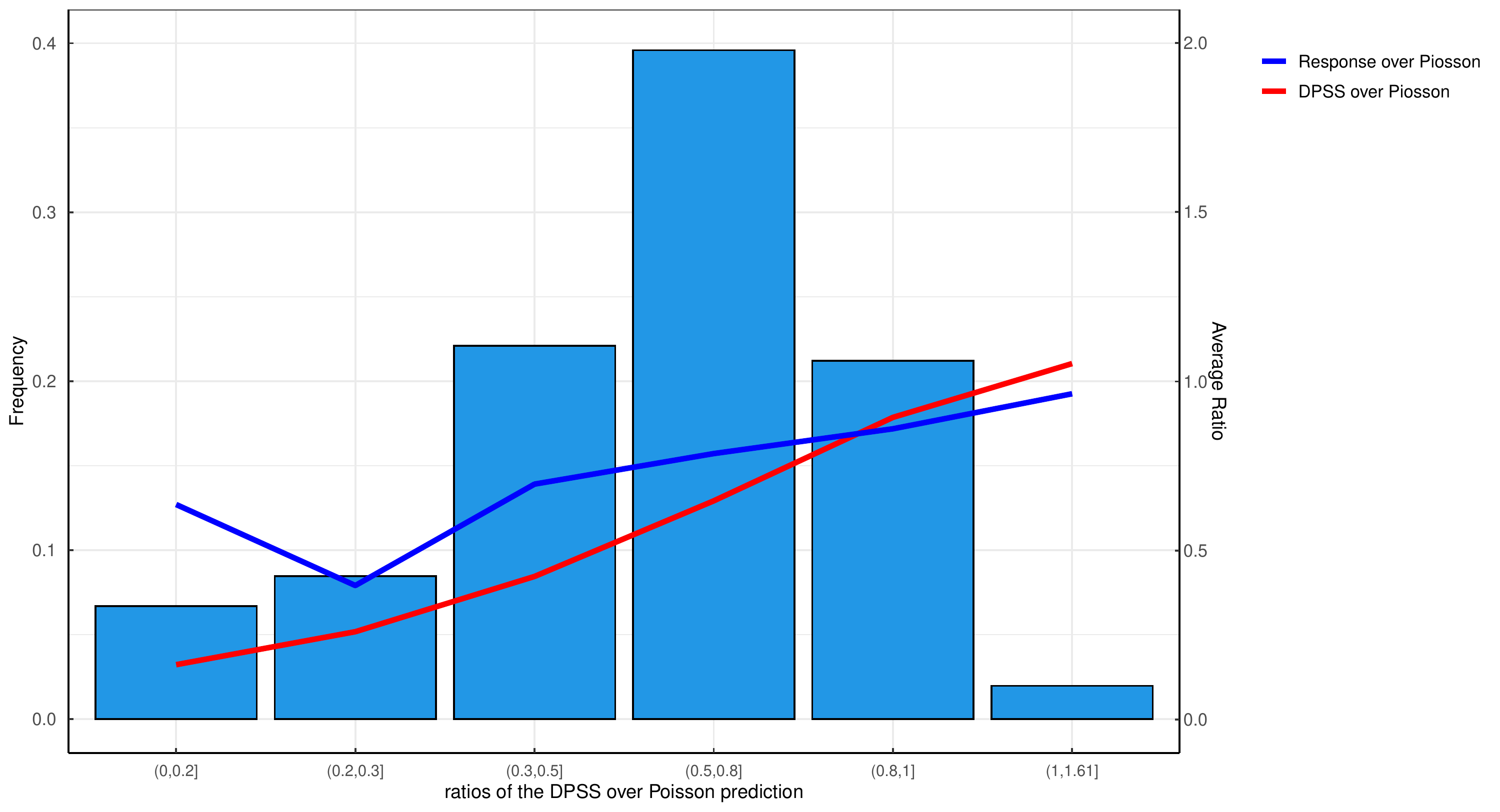}
	\end{center}
	\caption{Double lift plot for DPSS model over Poisson model.
		Observations
		are sorted in the ratios of the DPSS over the Poisson prediction and are grouped by the intervals that they belong (ratio of $0.6 \in (0.5, 0.8]$).
		For each bucket, the ratio of actual claim counts
		over total Poisson prediction is marked as the blue line, and
		the ratio of total DPSS prediction over total Poisson prediction is marked as the red line.
	}
	\label{fig-double-lift-plot}
\end{figure}

From Figure \ref{fig-lift-plot}, we can also assess the performance of the candidates by utilizing the lift plot \citep{lee2021addressing}.
To derive the measure, we sort
the prediction and group the observations into 10 deciles.
Figure \ref{fig-lift-plot} shows the average responses over average predictions for the deciles.
If the points are
aligned with 45 degree line, the model has a high predictive performance.
From Figure  \ref{fig-lift-plot}, we see that dynamic models perform extremely well in predicting high-frequency risk classes as the prediction and response
are almost the same.
Figure \ref{fig-double-lift-plot} shows the
double lift plot, which provides a direct predictive performance comparison between the DPSS model over the Poisson model.
The double lift plot is discussed in details in the work of \cite{Lee2019}.
We can see a significantly positive relationship between
the blue line (loss ration) and the red line (indicated rate change), with a correlation of 0.963,
indicating the DPSS model is able to explains a high portion of residuals that the Poisson model fails to capture.

\begin{table}
	\caption{The difference between predicted and observed claims for the testing data in the real data set.}
	\label{tab-diff-prediction}
	\begin{center}
    \begin{tabular}{cccc|ccc}
    	\toprule
    		\multirow{2}[0]{*}{	Count $(k)$} &  \multicolumn{3}{c}{Static models} &  \multicolumn{3}{c}{Dynamic models}\\
    		\cline{2-7}
    		& Poisson    & NB    & ZIP   & DPSS  & DNBSS & DZIPSS \\
	\hline
    0     & -1353 & -1058 & -1032 & 526   & 685   & -222 \\
1     & 1342  & 828   & 838   & -309  & -555  & 268 \\
2     & 27    & 172   & 172   & -181  & -113  & -25 \\
3     & -12   & 44    & 21    & -31   & -15   & -16 \\
4     & -3    & 11    & 1     & -5    & -2    & -4 \\
5     & -1    & 2     & 0     & -1    & -1    & -1 \\
6     & 0     & 1     & 0     & 0     & 0     & 0 \\
%	\hline
%	Total & 1344.4 & 1362.8 & 1247.1 & -786.5 & -834.7 & 1251.9 \\
	\bottomrule
\end{tabular}%
	\end{center}
\end{table}

\begin{table}
	\caption{The Poisson deviance, lifts and Gini index for the testing data in the real data set.}
	\label{tab-dev-gini}
	\begin{center}
    \begin{tabular}{cccc|ccc}
    	\toprule
	\multirow{2}[0]{*}{Models} & \multicolumn{3}{c}{Static models} & \multicolumn{3}{c}{Dynamic models} \\
		\cline{2-7}
	& Poisson & NB    & ZIP   & DPSS  & DNBSS & DZIPSS \\
	\hline
	Poisson Deviance & 0.300 & 0.300 & 0.299 & 0.299 & 0.300 & 0.296 \\
%	Two-way Lift (10\%) & 60.585 & 60.252 & 60.363 & 55.958 & 55.717 & 49.933 \\
%	Two-way Lift (20\%) & 21.811 & 21.405 & 21.445 & 21.353 & 21.544 & 22.886 \\
%	Two-way Lift (30\%) & 12.780 & 12.324 & 12.634 & 13.302 & 13.244 & 12.297 \\
%	Two-way Lift (40\%) & 7.303 & 7.602 & 7.408 & 7.376 & 7.446 & 8.077 \\
%	Two-way Lift (50\%) & 6.221 & 6.186 & 6.318 & 5.755 & 5.690 & 5.762 \\
%	Two-way Lift (60\%) & 5.210 & 5.148 & 5.109 & 4.650 & 4.630 & 5.403 \\
%	Two-way Lift (70\%) & 4.059 & 3.997 & 4.074 & 4.434 & 4.403 & 4.182 \\
%	Two-way Lift (80\%) & 2.749 & 2.734 & 2.744 & 3.050 & 3.066 & 2.927 \\
	Two-way Lift & 1.735 & 1.711 & 1.707 & 1.608 & 1.581 & 1.701 \\
	One-way Lift & 3.550 & 3.530 & 3.537 & 3.522 & 3.507 & 3.576 \\
	Gini index & 0.934 & 0.934 & 0.934 & 0.934 & 0.934 & 0.934 \\
%   Models       & Poisson & NB    & ZIP   & DPSS  & DNBSS & DZIPSS \\
%	\hline
%Poisson Deviance & 0.3003 & 0.3003 & 0.2995 & {0.2993} & 0.2995 & {0.2963} \\
%one-way Lift  & 3.5500 & 3.5305 & 3.5370 & 3.5218 & 3.5066 & 3.5760 \\
%Gini index & 0.9342 & 0.9342 & 0.9342 & {0.9343} & 0.9343 & 0.9342 \\
\bottomrule
\end{tabular}%
	\end{center}
\end{table}

%for capturing the differences between predictions and observations, due to the high proportions of zeros and the skewed heavy-tailed distribution of the positive losses.
%For this reason,
%we consider the ordered Lorenz curve and the associated Gini index as a statistical measure of the association between the premium and loss distributions in non-life insurance, through which different predictive models can be compared \citep{frees2011summarizing,yang2018insurance,shi2018pair}.

%\clearpage
\section{Conclusion}\label{section:conclusion}
We have proposed and studied the Dynamic Poisson state space model and its extensions to not only handle very unbalance claim data with excessive proportion of zeros, but update the prediction of claim frequency over time by the state space modelling framework.
Unlike Bayesian estimation methods used in state space models which are frequently discussed in the existing actuarial science literature \citep{Jae2022},
the parameters in our proposed model are modelled by the smoothing splines over time and are estimated via maximum likelihood method.
The prediction for claim frequency is based on Kalman filter like algorithm and can be updated dynamic when new claim information becomes available.
The proposed method overcomes the limitations of traditional varying-coefficients models which mainly focus on estimation and statistical inference rather than the future prediction.
It is also designed for online monitoring as the proposed algorithm has better computational efficiency than Bayesian estimation method.
In a real-world insurance claim data set, we model the claim counts data over six years as the function of several covariates in the Poisson, negative-binomial and zero-inflated Poisson assumptions and update the models every one year.
The empirical results show that the dynamic pattern of the estimated coefficients in DPSS for most of rating factors can be explained by the process of auto insurance rate reform in China from year 2010 to 2015.
Several practical metrics reveal the superiority of the proposed method in predicting claim frequency.

It is worth mentioning that the DPSS we proposed in this work were univariate count models,
and it would be extended to multivariate count models when the frequency of automobile claims contains more than one types, for example the
number of claims of third-party bodily injury ($Y_1$), the number of
claims of own damage ($Y_2$), and the number of claims of third party property damage ($Y_3$).
Another extension of this work could be devoted to the spatial analysis in the dynamic state space setting that allows us
to investigate more complex effects of the covariates on the insurance claim data.
Besides, the dynamic modelling approach based state space equations could be further  discussed for longitudinal claim data in the experience ratemaking context.

\section*{Supplementary materials}
{\textbf{Code}: The R source code for Dynamic Poisson state space model	 and simulation study. A readme file is included in the archive (dpss.zip).}

\section*{Acknowledgement}
Zhengxiao Li acknowledges the financial support from National Natural Science Fund of China (Grant No. 72271056),
the University of International Business and Economics project for Outstanding Young Scholars (Grant No. 20YQ16),
and ``the Fundamental Research Funds for the Central Universities" in UIBE (Grant No. CXTD13-02).
Jiakun Jiang acknowledges the financial support from National Natural Science Fund of China (Grant No. 12101054) and National Statistical Science Research Program (Grant No. 2022LY034).

\appendix
\section*{Appendices}
\addcontentsline{toc}{section}{Appendices}
\renewcommand{\thesubsection}{\Alph{subsection}}

\subsection{Dynamic zero-inflated Poisson state space model (DZPSS)}
\label{app-DZPSS}

Suppose the response variable $y_i$ is zero-inflated Poisson distribution with the mean $e_i\lambda_i$ and the zero-inflated probability $\phi_i$, that is
\begin{equation}
\label{eq-ZIP}
{{Y}_{i}}\sim\left\{ \begin{matrix}
&0,   &\text{with}\text{ probability} \quad {{\phi }_{i}},\\
&	\text{Poisson}\left( {{\lambda }_{i}} \right),  &\text{with} \text{ probability } 1-{{\phi }_{i}} ,\\
\end{matrix} \right.
\end{equation}
%The ZIP model can be viewed as a mixture of a point mass at zero
%and a Poisson distribution with mean $e_i \lambda_i$, where $e_i$ is the risk exposure.
In \eqref{eq-ZIP}, $0  \leq \phi_i  <1$, so
extra zeros in the data can be explicitly modelled. The additional
point mass at 0 allows ZIP model to produce more responses with
value 0 than those predicted by the Poisson distribution, and thus
accommodates the zero-inflated phenomenon.

The effects of covariates with time-varying and time-invariant coefficients
can be modelled as
\begin{align}
\log \left( {{\lambda }_{i}} \right)&=\bm{B}_{i,q_1}^{T}\bm{\beta} \left( {{t}_{i}} \right)+\bm{B}_{i,q_2}^{T}\bm{\varphi} ={{\bm{B}}^T_{i}}\bm{\gamma} \left( {{t}_{i}} \right),\nonumber \\
\text{logit}\left( {{\phi }_{{i}}} \right)&=\log\left(\frac{\phi_i}{1-\phi_i}\right)=\bm{G}_{i,q_1}^{T}\bm{\rho} \left( {{t}_{i}} \right)+\bm{G}_{i,q_2}^{T}{{\bm{\alpha} }}={{\bm{G}}^T_{i}}\bm{\gamma} \left( {{t}_{i}} \right),
\label{eq-DZPSM-1}
\end{align}
where ${{\bm{B}}_{i,q_1}}={{\left( 1,{{B}_{i,1}},\cdots,{{B}_{i,q_1}} \right)}^{T}}$ and ${{\bm{G}}_{i,q_1}}={{\left(1, {{G}_{i,1}},\cdots,{{G}_{i,q_1}} \right)}^{T}}$
are the covariates with time varying  coefficients $\bm{\beta} \left( t \right)={{\left( {{\beta }_{0}}\left( t \right),\cdots,{{\beta }_{q_1}}\left( t \right) \right)}^{T}}$ and $\bm{\rho} \left( t \right)={{\left( {{\rho }_{0}}\left( t \right),,{{\rho }_{q_1}}\left( t \right) \right)}^{T}}$.
${{\bm{B}}_{i,q_2}}={{\left( 1,{{B}_{i,1}},\cdots,{{B}_{i,q_2}} \right)}^{T}}$ and ${{\bm{G}}_{i,q_2}}={{\left(1, {{G}_{i,1}},\cdots,{{G}_{i,q_2}} \right)}^{T}}$ are the covariates with time-invariant coefficients $\bm{\varphi} ={{\left( {{\varphi }_{1}},\cdots,{{\varphi }_{q_2}} \right)}^{T}}$ and $\bm{\alpha} ={{\left( {{\alpha }_{1}},\cdots,{{\alpha }_{q_2}} \right)}^{T}}$.
${{\bm{B}}_{i}}={{\left( {\bm{B}}_{i,q_1}^{T},{\bm{B}}_{i,q_2}^{T},0,\cdots,0 \right)}^{T}}$ and ${{\bm{G}}_{i}}={{\left( 0,\cdots,0,G_{i,q_1}^{T},G_{i,q_2}^{T} \right)}^{T}}$ are the $2(q_1+q_2+1)$-dimensional vector of covariates with vector of coefficients $\bm{\gamma} \left( t \right)={{\left( \bm{\beta} {{\left( t \right)}^{T}},{\bm{\varphi }^{T}},\bm{\rho} {{\left( t \right)}^{T}},{\bm{\alpha }^{T}} \right)}^{T}}$.

The parameters can be estimated through maximizing
the following penalized log-likelihood function:
\begin{align*}
{{L}_{c}}\left( \bm{\gamma} ; \bm{y} \right)&=\sum\limits_{{{y}_{i}}=0}{\log \left\{ {{e}^{{{\bm{G}}^T_{i}}\bm{\gamma} \left( {{t}_{i}} \right)}}+\exp \left[ -{{e}^{{{\bm{B}}^T_{i}}\bm{\gamma} \left( {{t}_{i}} \right)}} \right] \right\}}+\sum\limits_{{{y}_{i}}>0}{\left[ {{y}_{i}}{{\bm{B}}^T_{i}}\bm{\gamma} \left( {{t}_{i}} \right) -{{e}^{{{\bm{B}}^T_{i}}\bm{\gamma} \left( {{t}_{i}} \right)}}-\log \left( {{y}_{i}}! \right) \right]}\\
&\quad -\sum\limits_{i=1}^{n}{\log \left[ 1+{{e}^{{{\bm{G}}^T_{i}}\bm{\gamma} \left( {{t}_{i}} \right)}} \right]}
- \frac{1}{2}\sum\nolimits_{j=0}^{q_1}{{{\tau }_{1j}}\int{{{\left\{ {{\bm{\beta} }_{j}}^{''}\left( t \right) \right\}}^{2}}dt
-\frac{1}{2}\sum\nolimits_{j=0}^{q_1}{\tau }_{2j}\int{\left\{ {{\bm{\rho}}_{j}}^{''}\left( t \right) \right\}}^{2}dt,}}
\end{align*}
%where $\gamma_j(t)$ denotes the $j$th component of the vector $\bm{\gamma}(t)$, and the $\gamma_j'(t)$
%is the its first derivative.

Similar to the DPSS proposed in \eqref{eq-DPSM-1},
the DZIPSS \eqref{eq-DZPSM-1} can be also represented in
a state space representation as
\begin{align}
\log \left( {{\lambda }_{i}} \right)&={{\bm{U}}^T_{{i}}}\bm{\zeta}\left({{t}_{i}} \right), \nonumber \\
\text{logit}\left( {{\phi }_{i}} \right)&={{\bm{V}}^T_{{i}}}\bm{\zeta}\left({{t}_{i}} \right), \nonumber \\
\bm{\zeta}\left( {{t}_{i}} \right)&={{\bm{T}}_{{i}}}\bm{\zeta}\left( {{t}_{i-1}} \right)+\bm{\eta}\left( {{t}_{i}},{{t}_{i-1}} \right), \quad \bm{\eta}\left( {{t}_{i}},{{t}_{i-1}} \right) \sim N\left( \bm{0,}{{\bm{Q}}_{i}} \right),
\label{eq-DZPSM-2}
\end{align}
where
\begin{align*}
&{\bm{U}_{i}}={{\left( 1,0,{{B}_{i1}},0,{{B}_{i2}},\cdots ,{{B}_{i{{q_1}}}},0,{{B}_{i{{q_1+1}}}},{{B}_{i{{q_1+2}}}},\cdots,{{B}_{i{{q_1+q_2}}}},0,0,\cdots,0\right)}^{T}},\\
& {\bm{V}_{i}}={{\left( 0,0,\cdots,0,1,0,{{G}_{i1}},0,{{G}_{i2}},\cdots ,{{G}_{i{{q_1}}}},0,{{G}_{i{{q_1+1}}}},{{G}_{i{{q_1+2}}}},\cdots,{{G}_{i{{q_1+q_2}}}} \right)}^{T}}, \\
&\bm{\zeta}\left( t \right)={{\left( {{\beta }_{0}}\left( {{t}_{i}} \right),{{\beta }_{0}}'\left( {{t}_{i}} \right),\cdots,{{\beta }_{q_1}}\left( {{t}_{i}} \right),{{\beta }_{q_1}}'\left( {{t}_{i}} \right),{\bm{\varphi }^{T}},{{\rho }_{0}}\left( {{t}_{i}} \right),{{\rho }_{0}}'\left( {{t}_{i}} \right),\cdots,{{\rho }_{q_2}}\left( {{t}_{i}} \right),{{\rho }_{q_2}}'\left( {{t}_{i}} \right),{\bm{\alpha }^{T}} \right)}^{T}},\\
%&\bm{\eta}\left( {{t}_{i}},{{t}_{i-1}} \right)={{\left( U_{10}^{T}\left( {{t}_{i}},{{t}_{i-1}} \right),...,U_{1p1}^{T}\left( {{t}_{i}},{{t}_{i-1}} \right),U_{20}^{T}\left( {{t}_{i}},{{t}_{i-1}} \right),...,U_{2q_1}^{T}\left( {{t}_{i}},{{t}_{i-1}} \right) \right)}^{T}},\\
&{{\bm{T}}_{i}}=\text{diag}(\overbrace{{{T}_{i}},\cdots,{{T}_{i}}}^{{{q}_{1}}+1},\overbrace{1,\cdots,1}^{{{q}_{2}}},\overbrace{{{T}_{i}},\cdots,{{T}_{i}}}^{{{q}_{1}}+1},\overbrace{1,\cdots,1}^{{{q}_{2}}}),\\
&{{\bm{Q}}_{\text{i}}}=\text{diag}(\overbrace{\tau _{10}^{-1}{{Q}_{i}},\cdots,\tau _{1{{q}_{1}}}^{-1}{{Q}_{i}}}^{{{q}_{1}}+1},\overbrace{0,\cdots,0}^{{{q}_{2}}},\overbrace{\tau _{20}^{-1}{{Q}_{i}},\cdots,\tau _{2{{q}_{1}}}^{-1}{{Q}_{i}}}^{{{q}_{1}}+1},\overbrace{0,\cdots,0}^{{{q}_{2}}}).
\end{align*}
The estimation and prediction algorithm is similar to model (\ref{eq-DPSM-2}).

\subsection{Dynamic negative-binomial state space model (DNBPSS)}\label{app-DNBPSS}
The negative-binomial model can arise from a two-stage model for the distribution of a discrete variable $Y$. We suppose there is an unobserved random variable $\Theta$ have a Gamma distribution $\Theta\sim\text{Gamma}(\alpha, \alpha)$ with mean $1$ and variance $1/\alpha$.
Then the model postulates that conditionally on $\Theta$, $Y$ is Poisson with mean $\mu \Theta$.
Then the marginal distribution of $Y$ is the negative-binomial distribution with mean $\mu$ and variance $\mu+1/\alpha \mu^2$.
The effects of covariates with time-varying and time-invariant coefficients in the Dynamic negative-binomial state space model
can be introduced in the mean $\mu$ of the negative-binomial distribution, that is
%be modelled as
%The state space model is
%$E$ having a gamma distribution gamma($\theta$)/$\theta$, that is with mean 1 and variance $1/\theta$. Then the model postulates that conditionally on $E$, $Y$ is Poisson with mean $\mu E$ and $\theta E \sim$ gamma($\theta$). Then the marginal distribution of $Y$ is the negative-binomial distribution. The state space model is
\begin{align}
\log \left(\mu_i \right)&=\bm{z}_{i}^{T}\bm{\gamma} \left( {{t}_{i}} \right),\ \ \ \ i=1,\cdots ,n  \nonumber\\
\bm{\gamma} \left( {{t}_{i}} \right)&={\bm{T}_{i}}\bm{\gamma} \left( {{t}_{i}} \right)+\bm{\eta} \left( {{t}_{i}},{{t}_{i-1}} \right),\ \ \bm{\eta} \left( {{t}_{i}},{{t}_{i-1}} \right)\sim N\left(\bm{0},{\bm{Q}_{i}} \right) ,
\end{align}
where
\begin{align*}
{\bm{z}_{i}}&={{\left( 1,0,{{x}_{i1}},0,{{x}_{i2}},\cdots ,{{x}_{i{{q}_{1}}}},0,{{x}_{i,{{q}_{1}}+1}},\cdots ,{{x}_{iq}} \right)}^{T}}, \\
\bm{\gamma} \left( t \right)&={{\left( {{\beta }_{0}}\left( t \right),{{{{\beta }'}}_{0}}\left( t \right),{{\beta }_{1}}\left( t \right),{{{{\beta }'}}_{1}}\left( t \right),\cdots ,{{\beta }_{{{q}_{1}}}}\left( t \right),{{{{\beta }'}}_{{{q}_{1}}}}\left( t \right),{{\alpha }_{1}},\cdots ,{{\alpha }_{{{q}_{2}}}} \right)}^{T}}, \\
\bm{\eta} \left( {{t}_{i}},{{t}_{i-1}} \right)&={{\left( U_{0}^{T}\left( {{t}_{i}},{{t}_{i-1}} \right),\cdots ,U_{{{q}_{1}}}^{T}\left( {{t}_{i}},{{t}_{i-1}} \right) \right)}^{T}},\\
{\bm{T}}&=\text{diag}\left( \overbrace{{{T}},\cdots ,{{T}}}^{{{q}_{1}}+1},\overbrace{1,\cdots ,1}^{{{q}_{2}}} \right),
{\bm{Q}}=\text{diag}\left( \overbrace{\lambda _{0}^{-1}{{Q}},\cdots ,\lambda _{{{q}_{1}}}^{-1}{{Q}}}^{{{q}_{1}}+1},\overbrace{0,\cdots ,0}^{{{q}_{2}}} \right).
\end{align*}

\bibliographystyle{plainnat}
\bibliography{mybibfile}
\end{document}